\documentclass{jpp}

\usepackage{newtxmath}

\usepackage[T1]{fontenc}

\usepackage{subfig}
\usepackage{csquotes}
\usepackage{xcolor}
\usepackage{physics}
\usepackage{natbib}
\usepackage{xpatch}

\usepackage{hyperref}
\hypersetup{
	colorlinks=true,
	breaklinks=true,
	allcolors=blue,
	frenchlinks=true
}

\makeatletter
\xpatchcmd\NAT@citex
{%
	\@citea\NAT@hyper@{%
		\NAT@nmfmt{\NAT@nm}%
		\hyper@natlinkbreak{\NAT@aysep\NAT@spacechar}{\@citeb\@extra@b@citeb}%
		\NAT@date
	}%
}
{%
	\@citea
	\NAT@nmfmt{\NAT@nm}%
	\NAT@aysep\NAT@spacechar
	\NAT@hyper@{\NAT@date}%
}
{}{}
\xpatchcmd\NAT@citex
{%
	\@citea\NAT@hyper@{%
		\NAT@nmfmt{\NAT@nm}%
		\hyper@natlinkbreak{\NAT@spacechar\NAT@@open\if*#1*\else#1\NAT@spacechar\fi}%
		{\@citeb\@extra@b@citeb}%
		\NAT@date
	}%
}
{
	\@citea
	\NAT@nmfmt{\NAT@nm}%
	\NAT@spacechar\NAT@@open\if*#1*\else#1\NAT@spacechar\fi
	\NAT@hyper@{\NAT@date}%
}
{}{}
\makeatother

\graphicspath{{./Figs/}}

\usepackage{bbold}

\usepackage[normalem]{ulem} 

\newcommand{\onehalf}{\ensuremath{{\frac{1}{2}}}} 

\newcommand{\kB}{\ensuremath{k_\mathrm{B}}}
\newcommand{\ci}{\mathrm{i}} 
\newcommand{\mat}[1]{\ensuremath{\mathsfbi{#1}}}
\renewcommand{\vec}[1]{\boldsymbol{#1} }

\newcommand{\btimes}{\boldsymbol{\times}}

\begin{document}
\title[Fluid-particle-in-cell method with Landau closures]
{Coupling multi-fluid dynamics equipped with Landau closures to the particle-in-cell method}
\shorttitle{Fluid-particle-in-cell method with Landau closures}

\shortauthor{Lemmerz et al.}
\author{Rouven Lemmerz \aff{1,2} \corresp{\email{rlemmerz@aip.de}},
	Mohamad Shalaby \aff{1}\corresp{\email{mshalaby@live.ca}}, 
	Timon Thomas \aff{1} \and
	Christoph Pfrommer\aff{1}}
\affiliation{\aff{1} Leibniz-Institut f{\"u}r Astrophysik Potsdam (AIP), An der Sternwarte 16, 14482 Potsdam, Germany \aff{2} University of Potsdam, Institute of Physics and Astronomy,
	Karl-Liebknecht-Str. 24-25, 14476 Potsdam, Germany}

\date{Accepted XXX. Received YYY; in original form ZZZ}
\maketitle

\begin{abstract}
The particle-in-cell (PIC) method is successfully used to study magnetized
plasmas. However, this requires large computational costs and limits simulations
to short physical run-times and often to setups in less than three spatial
dimensions. Traditionally, this is circumvented either via hybrid-PIC methods
(adopting massless electrons) or via magneto-hydrodynamic-PIC methods (modelling
the background plasma as a single charge-neutral magneto-hydrodynamical
fluid). Because both methods preclude modelling important plasma-kinetic
effects, we introduce a new fluid-PIC code that couples a fully explicit and
charge-conservative multi-fluid solver to the PIC code SHARP through a
current-coupling scheme and solve the full set of Maxwell's equations. This
avoids simplifications typically adopted for Ohm's Law and enables us to fully
resolve the electron temporal and spatial scales while retaining the versatility
of initializing any number of ion, electron, or neutral species with arbitrary
velocity distributions. The fluid solver includes closures emulating Landau
damping so that we can account for this important kinetic process in our fluid
species. Our fluid-PIC code is second-order accurate in space and time. The code
is successfully validated against several test problems, including the stability
and accuracy of shocks and the dispersion relation and damping rates of waves in
unmagnetized and magnetized plasmas. It also matches growth rates and saturation
levels of the gyro-scale and intermediate-scale instabilities driven by drifting
charged particles in magnetized thermal background plasmas in comparison to
linear theory and PIC simulations. This new fluid-SHARP code is specially
designed for studying high-energy cosmic rays interacting with thermal plasmas
over macroscopic timescales.
\end{abstract}



\section{Introduction}

Astrophysical plasmas naturally partition into thermal and non-thermal particle populations.
Provided particles collide frequently via (Coulomb) collisions, this eventually leads to a characteristic thermal Maxwellian phase-space distribution. This population can be reliably described with the fluid approximation, which characterizes a vast amount of particles by a few macroscopic fields in space (e.g., number density, mean velocity and temperature).
By contrast, the non-thermal cosmic ray (CR) ion population at energies exceeding GeV is mostly collisionless and interacts with the background plasma via wave-particle interactions, thus retaining its initial power-law distribution for much longer times \citep{blandfordParticleAccelerationAstrophysical1987, drainePhysicsInterstellarIntergalactic2011, zweibelBasisCosmicRay2017}. Low-energy CRs ($\lesssim$~GeV) more frequently experience Coulomb/ionisation collisions and as such have a direct influence on gas dynamics and molecular chemistry \citep{dalgarnoInterstellarChemistrySpecial2006, padovaniImpactLowEnergyCosmic2020}. CRs can excite and grow plasma waves via instabilities at which they scatter in pitch angle (i.e., the angle between momentum and magnetic field vector), thereby regulating their macroscopic transport speed and exchanging energy and momentum with the thermal population. Modelling these plasma processes requires to move beyond the classical fluid approximation.

During the process of diffusive shock acceleration, CRs stream ahead of the shock into the precursor region and drive non-resonant Alfv\'en waves unstable by means of their powerful current \citep{bellTurbulentAmplificationMagnetic2004,riquelmeNonlinearStudyBell2009,caprioliSimulationsIonAcceleration2014a}, which provides efficient means of increasing their wave-particle scattering and reducing the CR diffusion coefficient \citep{caprioliSimulationsIonAcceleration2014b}.
Upon escaping from the acceleration site into the ambient medium, CRs continue to drive Alfv\'en-waves through resonant instabilities. Scattering off of these self-induced waves regulates their transport speed \citep{kulsrudEffectWaveParticleInteractions1969,marcowithCosmicRaydrivenStreaming2021b,shalabyNewCosmicRaydriven2020}, which is determined by the balancing instability growth and wave damping \citep{thomasCosmicrayHydrodynamicsAlfv2019,thomasProbingCosmicRay2020}. In the interstellar medium, CRs provide a comparable if not dominant pressure, despite their negligible number densities in comparison to the thermal population, which makes them dynamically important \citep{boularesGalacticHydrostaticEquilibrium1990, drainePhysicsInterstellarIntergalactic2011}.
Their pressure gradient can drive outflows from the interstellar medium \citep{simpsonRoleCosmicRayPressure2016,girichidisCoolerSmootherImpact2018,farberImpactCosmicRay2018} so that powerful global winds emerge from galaxies \citep{uhligGalacticWindsDriven2012,hanasz2013,pakmorGalacticWindsDriven2016,ruszkowski2017} that enrich the circumgalactic medium in galaxy haloes with CRs that can also dominate the pressure support and modify the cosmic accretion of gas onto galaxies \citep{buckEffectsCosmicRays2020,jiPropertiesCircumgalacticMedium2020}. The degree to which CRs regulate galaxy formation critically depends on the efficiency of wave-particle interactions, which in turn depends on the amplitude of self-excited plasma waves \citep{thomasCosmicRaydrivenGalactic2022}. On even larger scales, CRs energised in jets of active galactic nuclei stream into the surrounding intracluster medium of cool core clusters and heat it via the excitation of Alfv\'en waves and the successive damping \citep{guo2008,pfrommer2013,ruszkowskiCosmicRayFeedbackHeating2017,jacobCosmicRayHeating2017}.
Because the plasma physics underlying these processes is highly non-linear, numerical calculations are needed to study these effects.

Due to its ability to resolve kinetic processes, the PIC method \citep{dawsonOneDimensionalPlasmaModel1962,langdonTheoryPlasmaSimulation1970,hockneyComputerSimulationUsing1988,birdsallPlasmaPhysicsComputer1991} has become one of the most used methods for studying plasmas from laboratory to astrophysical scales. 
Examples of that include revolutionizing our understanding of the rich physics found in collisionless shocks  \citep{spitkovskyParticleAccelerationRelativistic2008,Marcowith_2016-shock-review}, magnetic reconnection  \citep{Daughton2006,Daughton2011,sironiRelativisticReconnectionEfficient2014}, instabilities driven by highly relativistic electron-positron beams~\citep{Bret2010,resolution-paper,shalabyGrowthBeamplasmaInstabilities2018,shalabyGrowthLongitudinalBeam2020}, as well as the transport of non-thermal particle populations like CRs \citep{holcombGrowthSaturationGyroresonant2019,shalabyNewCosmicRaydriven2020}.
However, the PIC method needs to advance numerous particles per cell each time step, and thus it is quick to reach its computational limit.
Even one-dimensional simulations usually only capture dynamics on very short physical times and
the extent to which two or three-dimensional simulations can be performed is very
limited.

The time interval between the inverse of the electron plasma frequency, $\omega_{\rm e}^{-1}$, (which is necessary to ensure the stability of the PIC algorithm) and that of the ion plasma frequency, $\omega_{\rm i}^{-1}$, depends on the ion-to-electron mass ratio, since $\omega_{\rm i}^{-1}/\omega_{\rm e}^{-1} = (m_{\rm i}/m_{\rm e})^{1/2}$, assuming charge neutrality, i.e.\ that the electron and
ion densities are equal.
Therefore, one frequently used trick to increase the computational efficiency in PIC simulations is to adopt a reduced ion-to-electron mass ratio to bridge the gap between the smallest timescale in the simulation and the larger timescale on which interesting physical processes occur.
However, this might lead to artificial suppression of physical effects~\citep{Bret2010mime,Hong+2012,Moreno+2018}, including instabilities with excitation conditions that depend on the mass ratio \citep{shalabyNewCosmicRaydriven2020,Shalaby2022}. This shows the need for a more efficient numerical method to complement the accurate results achieved by PIC simulations in order to enable simulations of realistic physics occurring on longer timescales.
One possible method consists in using the less expensive fluid approximation, which works particularly well for collisional systems where frequent particle collisions maintain a thermodynamic temperature but is less well motivated in weakly collisional or even collisionless astrophysical plasmas where it cannot accurately capture some important microphysical plasma processes.

Multiple methods have been devised that combine the computational advantages of a fluid code, while trying to maintain some of the physics accuracy provided by the PIC method.
 Hybrid-PIC codes
\citep{lipatovHybridMultiscaleSimulation2002, gargateDHybridMassivelyParallel2007}
treat electrons as a massless fluid and ions as particles. With the assumption of charge
neutrality and the Darwin approximation (i.e., neglecting the transverse displacement current), these codes are able to overcome some computational barriers while
omitting effects on the electron time and length scale. 
Since this eliminates the need to resolve electron scales, the increase in
computational efficiency from pure-PIC to hybrid-PIC methods is roughly a factor
of $\left(m_\mathrm{i}/m_\mathrm{e}\right)^{1/2}$ in timescale and about the same factor in spatial scales.
On the other hand, an even more efficient method exists, that combines the magneto-hydrodynamic (MHD) description of the thermal background plasma with PIC methods to model the evolution of energetic particles such as CRs \citep{baiMAGNETOHYDRODYNAMICPARTICLEINCELLMETHODCOUPLING2015, vanMarle2018}, called MHD-PIC.
However, this method inherits the assumptions of MHD, in particular, the use of (simplified) Ohm's law by fully neglecting the displacement current, which precludes physics associated with higher-order terms of Ohm's law as well as the electron
dynamics.

In this paper we present a self-consistent algorithm that is suitable for simulating microphysical effects of CR physics by only applying the fluid approximation to thermal particles and solving the full set of Maxwell's equations. 
Our goal of this novel fluid-PIC method is to sacrifice as little physics accuracy as possible, while at the same time alleviating computational restraints by orders of magnitude for setups involving CRs (or similar, low density non-thermal particle populations interacting with a thermal plasma).
The fluid-PIC method, in essence, couples a multi-fluid solver to the PIC method by summing their contributions to the charge and current densities used to solve Maxwell's equations, and the resulting electromagnetic fields. Thus, the subsequent dynamics is dictated by fluid and PIC species. This enables treating any arbitrary number of species in thermal equilibrium by modelling them as separate fluids that interact electromagnetically with each other and with particles of arbitrary momentum distribution (modelled using the PIC method).
In contrast to MHD-PIC and hybrid-PIC methods, we do not explicitly assume Ohm's law, and instead, solve Maxwell's equations in a fully self-consistent manner in our fluid-PIC code. Therefore, displacement currents are included in our model and fast changes in the electric field and electron dynamics are captured. This, in turn, allows studying the interaction of high energy particles with the background plasma, e.g.\ to investigate CR streaming.
Another hybrid approach resolving electron timescales fully, but using pressure coupling, has been used for simulation of pick-up ions in the heliosphere by \citet{burrowsNewHybridMethod2014}.

Often implicit and semi-implicit methods are utilized for stability and resolution reasons to couple the multi-fluid equations to Maxwell's equations \citep{hakimHighResolutionWave2006, shumlakAdvancedPhysicsCalculations2011, wangExactLocallyImplicit2020}. However, this creates an interdependency between all fluids and has limited utility when coupled to explicit particles.
We have developed an explicit multi-fluid solver in which each fluid and particle species is agnostic about each other and the coupling is achieved via an indirect current-coupling scheme.
Because the PIC part of the code is the most computationally expensive part of the fluid-PIC, hybrid-PIC, and MHD-PIC methods, the computational efficiency is mostly determined by the number of particles required as well as the smallest time and length scales that need to be resolved. Hence, this fluid-PIC approach results in large speed-ups for CR propagation simulations in comparison to traditional hybrid-PIC codes, which treat every ion as a particle and need to initialise a large number of particles according to the density ratio, as well as in comparison to PIC-only simulations. 
Especially studying comic ray propagation in the interstellar medium, where the typical CR density is of the order $10^{-9}$  times the interstellar medium number density, is challenging. 
Since the fluid-PIC algorithm is faster by orders of magnitude in comparison to PIC in such a case, we can reach further into the realistic parameter regime without sacrificing some essential microphysics. 

One of the most important kinetic effects is arguably Landau damping. The fluid
description can emulate this effect using Landau closures
\citep{hammettFluidMomentModels1990, umanskyModelingTokamakDivertor2015, hunanaIntroductoryGuideFluid2019}, which necessitates the computation of the heat flux in Fourier space.
While Fourier transforms in 1D are not easily parallelizable, this bottleneck can partially be mitigated by performing global communications of the message-passing interface (MPI) in the background while processing the high computational load (e.g.\ resulting from evolving orbits of PIC particles) in the foreground. Simulations with periodic boundary conditions are currently handled by convolution with a finite-impulse-response (FIR) filter in our code, but other options are available in the literature \citep{dimitsFastNonFourierMethod2014, wangLandaufluidClosureArbitrary2019}. A number of simplifying local approximations exist as well \citep{wangComparisonMultifluidMoment2015, allmann-rahnTemperatureGradientDriven2018, ngImprovedTenmomentClosure2020}, which scale computationally well but become inaccurate for studying some multiscale plasma physics problems. Our code implements these different approaches so that an appropriate one can be chosen, dependent on the requirements of a simulation. Our implementation is massively parallelized and can be efficiently run on thousands of cores.
Furthermore, the fluid-PIC method allows for any multi-fluid setup.
As such, this framework allows for some straightforward extensions. 
Potentially, this involves a setup with actively participating neutrals to incorporate ion-neutral damping into this method. 
To this end, the coupling between different fluids needs to be extended by a collision term, which is left as a future extension to the code.

The outline of this paper is as follows. In section~\ref{sec:method}, we introduce the pillars of this method and describe the PIC method, the fluid solver, how we couple both methods by means of electromagnetic fields, and describe various implementations of the Landau closure. In section~\ref{sec:results}, we show validation tests of the fluid solver (shock tube tests), linear waves in an ion-electron plasma, and the damping rate of Langmuir waves in a single-electron fluid with Landau closures. We then investigate the non-linear effects of two interacting Alfv\'en waves as well as cosmic-ray-driven instabilities, where fluid-PIC and PIC results are compared. We conclude in section~\ref{sec:conclusion}. Throughout this work, we use the SI system of units.

\section{Numerical Method}
\label{sec:method}
After a review of the kinetic description of a plasma in section~\ref{sec:kineticP}, we briefly introduce our PIC method in section~\ref{sec:PIC}.
The fluid description for plasmas and its assumptions are given in section~\ref{sec:fluid}.
The finite volume scheme we use to numerically solve the compressible Euler equations is described in section~\ref{sec:finitevolume}, while the electromagnetic interactions of the fluid are described in section~\ref{sec:fluidcharged}.
In section~\ref{sec:heatflux}, we describe the Landau closure we adopt in order to mimic the Landau damping in kinetic thermal plasmas within the fluid description, and detail its implementation in our code.
We close this section by describing the overall code structure of the fluid-PIC algorithm and finally discuss the interaction between the modules via the current-coupling scheme (Section~\ref{sec:fluid-PIC}).

\subsection{Kinetic description of a plasma}
\label{sec:kineticP}
The kinetic description of a collisionless relativistic plasma with particles of species $\mathrm{s}$ with
elementary mass, $m_\mathrm{s}$, and elementary charge, $q_\mathrm{s}$, is given by
the Vlasov equation,
\begin{equation}
\label{eq:Vlasov}
\pdv{f_\mathrm{s}}{t} + \frac{\vec{u}}{\gamma} \bcdot \vnabla f_\mathrm{s} + \vec{a}_\mathrm{s} \bcdot \vnabla_{u} f_\mathrm{s} = 0,
\end{equation}
where $f_\mathrm{s}=f_\mathrm{s}(\vec{x}, \vec{\varv},t)$ is the distribution function,
$\vec{u} = \gamma \vec{\varv}$ is the spatial component of the four-velocity
with the Lorentz factor $\gamma = [1 + \left(\vec{\varv}/c\right)^2]^{-1/2}$,
and $c$ is the light speed. The acceleration due to the Lorentz force is given
by
\begin{equation}
\vec{a}_\mathrm{s}  = \frac{q_\mathrm{s}}{m_\mathrm{s}} \left[ \vec{E}\left(\vec{x}, t\right) + \vec{\varv}  \cross \vec{B}\left(\vec{x}, t\right) \right],
\end{equation}
where $\vec{E}\left(\vec{x}, t\right)$ and $\vec{B}\left(\vec{x}, t\right)$ are
the electric and magnetic fields, respectively.
The evolution of electric and magnetic fields is governed by Maxwell's equations:
\begin{align}
\label{eq:MaxwellB}
\pdv{\vec{B}}{t} &= - \vnabla\cross\vec{E},
&\vnabla\bcdot\vec{B} &= 0,\\
\label{eq:MaxwellE}
\pdv{\vec{E}}{t} &= c^2 \vnabla\cross\vec{B}  - \frac{ \vec{J} }{ \varepsilon_0 } ,
&\vnabla\bcdot\vec{E}&= \frac{\rho}{\varepsilon_0},
\end{align}
where $c=1/\sqrt{\varepsilon_0 \mu_0}$ is the vacuum speed of light, and $\varepsilon_0$ and $\mu_0$ are the permittivity and the permeability of free space, respectively.
The evolution of the electro-magnetic fields is influenced by the charge density, $\rho$, and current density, $\vec{J}$.
They are given by the charge-weighted sum over all species of the number densities $n_\mathrm{s}$ and bulk velocities $\vec{\varw}_\mathrm{s}$ respectively,
\begin{alignat}{2}
\rho\left(\vec{x}, t\right)
  &= \sum_s q_s n_s \left(\vec{x}, t\right) &&= \sum_s q_s \int f_s\left(\vec{x}, \vec{\varv}, t\right) \mathrm{d}^3\varv,	  \label{eq:moments_rho}\\
\vec{J}\left(\vec{x}, t\right)
  &= \sum_s q_s  n_s \left(\vec{x}, t\right)  \vec{\varw}_s \left(\vec{x}, t\right) &&= \sum_s q_s \int  \vec{\varv} f_s\left(\vec{x}, \vec{\varv}, t\right) \mathrm{d}^3 \varv.
  \label{eq:moments_energy}
\end{alignat}

\subsection{The particle-in-cell method}
\label{sec:PIC}

We use the PIC method to solve for the evolution of plasma species that are modelled with the kinetic description.
The PIC method initializes a number of computational macroparticles to approximate the distribution function in a Lagrangian fashion.  Each macroparticle represents multiple physical particles and, as such, each macroparticle has a shape in position space which can be represented by a spline function. By depositing the particle motions and positions to the numerical grid (or computational cells), the electromagnetic fields can be computed.  This step is followed by a back-interpolation of these fields to the particle positions so that the Lorentz forces on the particles can be computed.
In our implementation, these equations are solved using one spatial
dimension and three velocity dimensions (1D3V), i.e.\ $\vnabla =
\left(\partial/\partial x, 0, 0\right)^\mathrm{T}$.

The code quantities are defined as multiples of the fiducial units given for time, fields (electric and magnetic), charge, current density and length
\begin{equation}
\begin{split}
t_0 = \sqrt{ m_0 \epsilon_0/( q_0^2 n_0) }, 
\hspace{0.5cm}
E _0   =  \sqrt{n_0 m_0 c^2/\epsilon_0},
\\ 
\rho_0 =  q_0 n_0,
\hspace{0.5cm}
J_0 =   \rho_0 c,
\hspace{0.5cm} 
x_0 = c t_0.
\end{split}
\label{eq:normalisations}
\end{equation}
This enables us to select a fixed time step of
\begin{equation}
\Delta t = C_\mathrm{cfl} c \Delta x
\label{eq:time stepPIC}
\end{equation}
where $C_\mathrm{cfl}< 0.5$ to satisfy the Courant-Friedrichs-Lewy (CFL) condition.
The value of the reference density $n_0$ is chosen such that the code timescale, $t_0$, obeys $\omega_\mathrm{p}^{-2}= t_0^2$. 
The total plasma frequency is $\omega_\mathrm{p}=(\sum_s \omega_s^2)^{1/2}$, and related to the plasma frequencies of the individual species, $\omega_\mathrm{s}^2 = q_\mathrm{s}^2 n_\mathrm{s}/(m_{s} \epsilon_0)$.
We define the discretized time $t^{k} = k \Delta t$, position $x_i = i \Delta x$
and quantities at discrete position and times as $\vec{E}^{k}_i = \vec{E}(t^{k},
x_i)$. For details on the PIC code SHARP, the reader is referred to
\citet{shalabySHARPSpatiallyHigherorder2017, shalabyNewCosmicRaydriven2020}.
Here, we focus on describing how SHARP is extended to include fluid treatment of some plasma species.

\subsection{Fluid description of plasma}
\label{sec:fluid}
A straightforward way of coarse graining the Vlasov equation
\eqref{eq:Vlasov} is to reduce its dimensionality.  
By taking the $j$-th moment over velocity space, i.e.\ $\int \vec\varv^j f \mathrm{d}^3\varv$, we retrieve the fluid quantities and reduce the dimensionality of the 1D3V kinetic description to 1D.
The number density $n_\mathrm{s}$ and the bulk velocity $\vec{\varw}_\mathrm{s}$ are defined through the zeroth and first moment of the distribution function, respectively, while the total energy density per unit mass $\epsilon_s$ and the scalar pressure per unit mass $p_s$ are related to the second moment \citep{wangComparisonMultifluidMoment2015}:
\begin{align}
	n_\mathrm{s} \left(\vec{x}, t\right)  &= \int  f_\mathrm{s}\left(\vec{x}, \vec{\varv}, t\right) \mathrm{d}^3 \varv,\\
	 \vec{\varw}_\mathrm{s} \left(\vec{x}, t\right)  &= \int\dfrac{1}{n_\mathrm{s}\left(\vec{x}, t\right) }  \vec{\varv} f_\mathrm{s}\left(\vec{x}, \vec{\varv}, t\right) \mathrm{d}^3 \varv ,\\
	\epsilon_\mathrm{s} \left(\vec{x}, t\right) &=\int  \dfrac{1}{2}  \vec\varv^2 f_\mathrm{s}\left(\vec{x}, \vec{\varv}, t\right) \mathrm{d}^3 \varv, \\
	p_\mathrm{s} \left(\vec{x}, t\right)  &=  \int  (\varv_x - \varw_{\mathrm{s},x})^2 f_\mathrm{s}\left(\vec{x}, \vec{\varv}, t\right) \mathrm{d}^3 \varv 
	= \frac{\Gamma - 1}{2}  \int  (\vec\varv - \vec\varw_\mathrm{s})^2 f_\mathrm{s}\left(\vec{x}, \vec{\varv}, t\right) \mathrm{d}^3 \varv. \label{eq:pressure_moment}
\end{align}
Here, the pressure tensor is under the adiabatic assumption  and the degrees of freedom are encoded in the adiabatic index $\Gamma$.
The following relation is found from the definitions
\begin{align}
	\epsilon_\mathrm{s} = \frac{p_\mathrm{s}}{\Gamma - 1} + \frac{1}{2}\,n_\mathrm{s} \,\vec{\varw}_\mathrm{s} \bcdot \vec{\varw}_\mathrm{s}.
\end{align}

The first three moments of the of the Vlasov equation are called the continuity, momentum, and energy conservation equations. A set of these equations is found for each fluid species, but the subscript $\mathrm{s}$ is neglected here for simplicity:
	\begin{align}
	\frac{\partial n}{\partial t} + \vnabla \bcdot \left(n  \vec{\varw}  \right) &= 0, \label{eq:mass_conservative}\\
	\frac{\partial n\vec{\varw}  }{\partial t} + \vnabla \bcdot \left[ p \mat{1} + n \vec{\varw} \vec{\varw} \right]&= \frac{q}{m} \vec{S}_\varw\left(n, \vec{\varw}, \vec{B}, \vec{E}\right), \label{eq:mom_conservative}\\
	\frac{\partial \epsilon}{\partial t} + \vnabla \bcdot \left[ (p  + \epsilon ) \vec{\varw}  \right] + \frac{1}{\Gamma - 1} \vnabla \bcdot \vec{Q}&= \frac{q}{m} \vec{\varw}\bcdot\vec{S}_\varw\left(n, \vec{\varw}, \vec{B}, \vec{E}\right).
	\label{eq:cons_energy}
	\end{align}
We assumed the non-relativistic limit and an isotropic pressure tensor with vanishing non-diagonal components, i.e.\ the inviscid limit.
The notation $\vec{\varw}\vec{\varw}$ indicates the dyadic product of the two
vectors and $\mat{1}$ is the unit matrix.
Similar to the definition of the scalar pressure in equation~\eqref{eq:pressure_moment} we use a definition of the heat flux vector, which is normalized to the degrees of freedom as well
\begin{equation}
\vec{Q} \left(\vec{x}, t\right)  = \frac{\Gamma - 1}{2} \int \left(\vec\varv-\vec\varw\right)^2 \left(\vec\varv - \vec\varw\right) f\left(\vec{x}, \vec{\varv}, t\right)  \mathrm{d}^3\varv . 
\end{equation}
The electromagnetic source term is given by
\begin{equation}
\vec{S}_\varw\left(n, \vec{\varw}, \vec{B}, \vec{E}\right) = n  \left(\vec{E} + \vec{\varw} \btimes \vec{B}\right).
\end{equation}

The general form of the fluid equations can be written as 
\begin{equation}
	\pdv{\tilde{\vec{U}}}{t} + \vnabla \bcdot\mat{F}(\tilde{\vec{U}}) = \vec{S}(\tilde{\vec{U}}),
	\label{eq:fluidstate}
\end{equation} 
where $\tilde{\vec{U}} = \tilde{\vec{U}}(\vec{x}, t) = \left(n, n\vec{\varw},
\epsilon\right)^\mathrm{T}$ is the fluid state vector at position $(\vec{x}, t)$, $\mat{F}$ is the flux matrix,
and $\vec{S}$ is the source vector. 

Numerically, the complexity of solving equation~\eqref{eq:fluidstate} can be reduced by splitting the operator into less complex sub-operators using Strang operator splitting \citep{strangConstructionComparisonDifference1968,hakimHighResolutionWave2006}. 
This enables us to use the most appropriate solver for each subsystem sequentially. 
We split the fluid update into three parts; the flux $\mat{F}$ excluding the heat flux (see section~\ref{sec:finitevolume}), the electromagnetic source $\vec{S}_{\rm em}=\vec{S}_\varw q/m$ (see section~\ref{sec:fluidsource}), and the heat flux $\vec{Q}$ (see section~\ref{sec:heatflux}). 
For commuting operators $\exp(\Delta t \vec{Q})$ and $\exp(\Delta t \vec{S}_{\rm em})$ a second order accurate Strang splitting is obtained as
\begin{align}
	\vec{U}^{n+\onehalf} &= \mathrm{e}^{\frac{\Delta t}{2} \mat{F}} \mathrm{e}^{\Delta t \vec{Q}}   \mathrm{e}^{\Delta t \vec{S}_{\rm em} } \mathrm{e}^{\frac{\Delta t}{2} \mat{F}} \vec{U}^{n-\onehalf} + O(\Delta t^3) .
	\label{eq:splitting}
\end{align}
If $\vec{Q}$ and $\vec{S}$ act independently on the entries $p$ and $\vec{\varw}$ respectively, then the order of applying them can be varied and they need to be evaluated only once.
In practice the formulation of $\vec{Q}$ might partially depend on $\vec{\varw}$. In this case, Strang splitting is performed on this part of the operator $\vec{Q}$ as well, see equation~\eqref{eq:splitting_detailed}.

\subsection{Finite volume scheme}
\label{sec:finitevolume}

The 1D3V fluid equations are solved using a finite volume method, where the
fluid equations are averaged over the cell volume, which is an interval of
length $\Delta x$ in 1D,
\begin{equation}
\label{eq:FVcell}
\vec{U}_i\left(t\right) = \frac{1}{\Delta x}\int_{x_{i-\onehalf}}^{x_{i+\onehalf}} \tilde{\vec{U}}\left(x, t\right)\mathrm{d}x.
\end{equation}
This enables us to correctly conserve the overall fluid mass, fluid momentum and
fluid energy, even in the presence of large gradients, by utilizing Gauss' theorem:
\begin{equation}
\frac{1}{\Delta x} \int_{x_{i-\onehalf}}^{x_{i+\onehalf}} \pdv{\vec{F}(\tilde{\vec{U}})}{x}\mathrm{d}x = \frac{1}{\Delta x} \left[ \vec{F}_{i+\onehalf} - \vec{F}_{i-\onehalf}\right]
\end{equation}
where the flux through an interface at $x_i$ is $\vec{F}_{i}\left(t\right) = \vec{F}[\tilde{\vec{U}}(x_{i}, t)]$, leading to the update equation
\begin{equation}
\label{eq:FVUpdate}
\pdv{\vec{U}_i(t)}{t} = \frac{1}{\Delta x} \left[ - \vec{F}_{i+\onehalf} + \vec{F}_{i-\onehalf} + \int \vec{S}\big(\tilde{\vec{U}}(x, t)\big) \mathrm{d}x \right].
\end{equation}
Integrating equation~\eqref{eq:FVUpdate} in time is achieved by using second, third, or fourth-order Runge-Kutte methods~\citep{butcherNumericalMethodsOrdinary2016}.
In contrast to the finite difference scheme used for electromagnetic fields and particles, where electromagnetic quantities are point values,
fluid quantities discretized with the finite volume method are cell averages.
This is useful, because the finite difference method does not guarantee the conservation of the conservation equations~\eqref{eq:mass_conservative} through~\eqref{eq:cons_energy}, which are governing the fluid; while on the other hand using the finite volume method for the electromagnetic fields needs additional steps to satisfy the  constraint $\vnabla\bcdot \vec{B}=0$.
Hybridization of both schemes to combine the advantages of each has been used before in other contexts, i.e.\ \citet{soaresfrazaoUndularBoresSecondary2002}.

The maximum time step in the 1D3V Euler equations, which allows for stable simulations, is $\Delta
t<C_\mathrm{cfl}\Delta x \times (\abs{\varw} + c_\mathrm{s})$, with the speed of sound $c_\mathrm{s}
= (\Gamma p/n)^{1/2}$. 
For all realistic setups these velocities are limited naturally by the speed of light, $\abs{\varw} < c$ and $c_\mathrm{s} < c$, and this condition is automatically fulfilled by the time step criterion in equation~\eqref{eq:time stepPIC}.
In practice, only equation~\eqref{eq:time stepPIC} together with a suitable Courant number of $C_\mathrm{cfl}\leq 0.5$ is used to determine the time step of the simulation.

\subsubsection{Reconstruction}
\label{sec:recon_cweno}
To approximate the flux at interfaces, we need to reconstruct the fluid state at cell interfaces. The accuracy of the reconstruction has a crucial
influence on the diffusivity. A lower-order reconstruction can lead to excessive damping of waves, which might suppress relevant physical effects on longer timescales.

For reconstructing the point value $\tilde{\vec{U}}(x_{i+{1/2}}, t)$,
which is needed to compute $\vec{F}_{i+{1/2}}$, we employ a central weighted essentially non-oscillatory reconstruction (C-WENO) scheme of spatial
order five.  
The reconstruction computes two point values at each
interface $x_{i+{1/2}}$, an interpolation from the left- and
right-hand side. 
We reconstruct the primitive variables $n$, $\vec{\varw}$, and $p$ individually.

Our implementation of the C-WENO method is based on the 5th order scheme presented in \citet{2008Capdeville}. An introduction to the topic can be found in \citet{2017Cravero}.
The C-WENO reconstruction uses a convex combination of multiple low-order reconstruction polynomials to achieve high-order interpolations of the interface values while it employs a non-linear limiter to degrade this high-order interpolation to a lower order if the reconstructed quantity contains discontinuities. The fifth-order C-WENO uses three third-order polynomials $P_\mathrm{L}(x), P_\mathrm{C}(x), P_\mathrm{R}(x)$ for each cell $i$ to interpolate the four adjacent cells in the following way:

\begin{tabular}{llccc}
$P_\mathrm{L}(x)$ &\text{interpolates values at} & $i - 2$ & $i - 1$ & $i$     \\
$P_\mathrm{C}(x)$ &\text{interpolates values at} & $i - 1$ & $i$     & $i + 1$ \\
$P_\mathrm{R}(x)$ &\text{interpolates values at} & $i   $  & $i + 1$ & $i + 2$\\
\end{tabular}

\noindent
while the optimal fifth-order polynomial interpolates all of them:

\begin{tabular*}{\columnwidth}{llccccc}
$P_\mathrm{opt}(x)$&interpolates values at&$i-2$&$i-1$&$i$&$i+1$&$ i+2$.\\
\end{tabular*}

We define an additional polynomial 
\begin{equation}
P_0(x) = \frac{1}{d_0} \left[P_\mathrm{opt}(x)  - \sum_{q \in [\mathrm{L, C, R}]} d_q P_q(x) \right], \label{eq:P_0_definition}
\end{equation} 
where $d_0 + d_\mathrm{L} + d_\mathrm{C} + d_\mathrm{R} = 1$. The polynomials $P_0$, $P_\mathrm{L}$, $P_\mathrm{C}$, and $P_\mathrm{R}$ are a convex representation of the $P_\mathrm{opt}$ polynomial. 
We use $d_0 = 3 / 4$, $d_\mathrm{C} = 2 / 16$, and $d_\mathrm{L} = d_\mathrm{R} = 1 /16$.

In general, we would like to use the reconstruction provided by the $P_\mathrm{opt}$ polynomial as frequently as possible because of its high-order nature. 
But this high-order reconstruction can cause oscillations similar to the Gibbs phenomenon at discontinuities.
Therefore, we need to employ a limiting strategy to avoid such behaviour. In order to accomplish this, we re-weight all of our $d$-coefficients by taking the smoothness of the associated polynomial into account \citep{1996Jiang}. We define
\begin{align}
\alpha_q = d_q \left[1 + \left(\frac{\tau}{\textsc{IS}[P_q] + 10^{-9} \Delta x }\right)^2 \right] \quad \text{for } q \in [\mathrm{0, L, C, R}],
\end{align}
where $\tau$ is a measure for the overall smoothness of the reconstructed variables, and $\mathrm{IS}[P_q]$ defines a smoothness indicator of the low-order polynomials.
Because the formulae for these smoothness indicators are quite cumbersome, we list them in appendix~\ref{sec:cweno_coefficients}.
These coefficients define a new set of normalized weights given by
\begin{equation}
 w_q = \frac{\alpha_q}{\alpha_0 + \alpha_\mathrm{L} + \alpha_\mathrm{C} + \alpha_\mathrm{R}} \quad \text{for } q \in [\mathrm{0, L, C, R}].
\end{equation}
The final reconstructed polynomial is then given by the convex combination of the low-order polynomials using this set of normalized weights:
\begin{align}
P_\mathrm{rec}(x) = w_0 P_0(x) + w_\mathrm{L} P_\mathrm{L}(x) + w_\mathrm{C} P_\mathrm{C}(x) + w_\mathrm{R} P_\mathrm{R}(x),
\end{align}
which we evaluate at the cell interfaces to calculate the required left- and right-handed interface values for the Riemann solver. We detail how these polynomials are evaluated in appendix~\ref{sec:cweno_coefficients}.

The smoothness indicators $\textsc{IS}[P_q]$ vanish if the underlying polynomials are smooth.
In this case, the re-weighted coefficients reduce to their original value $\alpha_q \rightarrow d_q$ and the reconstructed polynomial reduces to the optimal polynomial $P_\mathrm{rec}(x) \rightarrow P_\mathrm{opt}(x)$.


\subsubsection{Riemann solver}
\label{sec:riemann}
The previous reconstruction step determines two, potentially different, values $\tilde{\vec{U}}_\mathrm{L}$ and $\tilde{\vec{U}}_\mathrm{R}$ for each quantity to the left and right of every interface, thereby providing the initial conditions for the Riemann problem:
\begin{align}
\pdv{\vec{U}}{t} &= - \vnabla \bcdot\mat{F}(\tilde{\vec{U}}) \\
\tilde{\vec{U}}(x, 0) &= \begin{cases}
\tilde{\vec{U}}_\mathrm{L}, \quad x<0 \\
\tilde{\vec{U}}_\mathrm{R}, \quad x>0
\end{cases}
\end{align}
An (approximate) Riemann solver is employed to compute the numerical flux $\mat{F}(\tilde{\vec{U}})$.
While a number of different families of Riemann solvers have been developed with individual strengths and weaknesses, we have decided to implement multiple solvers which can be changed on demand.
Implemented solvers in fluid-SHARP include a Roe solver with entropy fix \citep{roeApproximateRiemannSolvers1981, hartenSelfAdjustingGrid1983} and an HLLC solver \citep{toroRestorationContactSurface1994}.
While the Roe solver yields more accurate solutions and fewer overshoots in our tests in comparison to the HLLC solver, it becomes unstable in near vacuum flows and strong expansion shock waves.
Even though differences between the solvers are easily visible in some shock setups and artificially extreme conditions, they are typically negligible in most applications common for thermal plasmas.
We opt to employ the HLLC solver as our standard for stability purposes and use the Roe solver in cases where stronger shocks with overshoots are expected.

\subsection{Electromagnetic interaction with charged fluids}
\label{sec:fluidcharged}

In this section, we first introduce the Lorentz force as a source term in equation~\eqref{eq:mom_conservative}. 
Furthermore, we describe how the fluid influences the electromagnetic fields.
With these two additional parts, the description from an uncharged gas in section~\ref{sec:finitevolume} is expanded here to include plasmas.
 
\subsubsection{Treatment of electromagnetic source term}
\label{sec:fluidsource}
Instead of integrating the energy equation~\eqref{eq:cons_energy}, which would require evaluating the source term on the right-hand side, we compute the time
evolution of the primitive pressure variable, for which the electromagnetic
source term conveniently vanishes:
\begin{equation}
\label{eq:pressure_update}
\frac{\partial p }{\partial t} + \Gamma p \vnabla\bcdot\vec{\varw} + \vec{\varw} \bcdot \vnabla p + \vnabla \bcdot \vec{Q} = 0.
\end{equation}

Then only the computation of the source term for the momentum
equation~\eqref{eq:mom_conservative} is left, which uses the Boris integrator \citep{boris1970relativistic} to account
for the Lorentz force on the fluid momentum vectors.
Up until now we have only applied the C-WENO method for conservation laws, however, by adding the source term, we are left with a balance law.
In C-WENO formulations for balance laws it is customary to approximate the integral of the source term (equation~\ref{eq:FVUpdate}) numerically to higher orders as well \citep{craveroCWENOUniformlyAccurate2018}.
We use Simpson's Formula for approximating equation~\eqref{eq:FVUpdate}
\begin{equation}
\int_{x_{i-1/2}}^{x_{i+1/2}} \vec{S}\left(\tilde{\vec{U}}\right) \mathrm{d}x = \frac{1}{6}\left(\vec{S}(\tilde{\vec{U}}_{i-\onehalf}) + 4 \vec{S}(\tilde{\vec{U}}_{i}) + \vec{S}(\tilde{\vec{U}}_{i+\onehalf})\right) + \textit{O}(\Delta x^5),
\end{equation}
where the intra-cell values $\tilde{\vec{U}}_{i\pm1/2}$ are interpolated by the same C-WENO scheme as used for solving the hydrodynamical equations, and the centre-value is computed self-consistently with the numerical integration formula, i.e.\ $\tilde{\vec{U}}_{i} = (6 \vec{U}_{i} - \tilde{\vec{U}}_{i+1/2} - \tilde{\vec{U}}_{i-1/2})/4$.
We also need to interpolate the electromagnetic field values to a comparable spatial order.
This is achieved by performing finite-difference interpolations for each component from the Yee mesh discretized fields, that is
\begin{align}
E_{i+\onehalf} &= \dfrac{150 (E_{i}+E_{i+1}) - 25 (E_{i-1}+E_{i+2}) + 3  (E_{i-2}+E_{i+3})}{256}
+ \textit{O}(\Delta x^6),
\end{align}
and temporal order, $B^n=(B^{n+1/2}+B^{n-1/2})/2$, again, for each component necessary.
Lower order approximations produce, in our tests, similar results, but converge to slightly lower wave frequencies when compared with the analytical solution of the dispersion relation.

\subsubsection{Deposition of charges}
Equations~\eqref{eq:MaxwellE} govern the electric field evolution, where Faraday's or Gauss' law might be used to compute $\vec{E}$.
In this section we focus on the one-dimensional setup without particle contributions, which are explained in section~\ref{sec:fluid-PIC}.
The perpendicular components' update, $E_y$ and $E_z$, is received straightforwardly by discretizing Faraday's Law
\begin{align}
\left(E_{y}\right)^{n+1}_{i+\onehalf} &= \left(E_{y}\right)^n_{i+\onehalf} -\sum_s \frac{\Delta t}{\epsilon_0} q_s \left(n \varw_y\right)_{i+\onehalf, s}^{n+\onehalf}
 - \frac{c^2 \Delta t}{\Delta x} \left[ \left(B_z\right)_{i+1}^{n+\onehalf} - \left(B_z\right)_{i}^{n+\onehalf} \right] \\
\left(E_{z}\right)^{n+1}_{i+\onehalf} &= \left(E_{z}\right)^n_{i+\onehalf} -\sum_s \frac{\Delta t}{\epsilon_0} q_s \left(n \varw_z\right)_{i+\onehalf, s}^{n+\onehalf} 
+ \frac{c^2\Delta t}{\Delta x} \left[ \left(B_y\right)_{i+1}^{n+\onehalf} - \left(B_y\right)_{i}^{n+\onehalf} \right] ,
\end{align}
where the sum is taken over all fluid species $\mathrm{s}$ and $n \vec{\varw}$ are components of the fluid vector $\vec{U}$.

For the $E_x$ component in spatial direction however, in order to enforce charge-conservation, Gauss' law in discretized form needs to be enforced for all $i\ge 1$ as well 
\begin{align}
\label{eq:ExGauss}
\left(E_{x}\right)^{n}_i = \left(E_{x}\right)^n_0 + \sum_s \frac{q_s}{\epsilon_0} \sum_{j=0}^{i-1}n^n_{j+\onehalf, s} \Delta x =\left(E_{x}\right)^n_0 + \sum_s \frac{q_s}{\epsilon_0} \int_{x_{0}}^{x_{i}} \tilde{n}_s^n \mathrm{d}x,
\end{align}
where the second equality uses the definition of cell averages in the finite volume scheme (see equation~\ref{eq:FVcell}) and shows, that this numerical formula is exact.
Another formula for updating $(E_x)_0$ to the time step $n$ is still needed.
In the analytical case Gauss' law in combination with the density conservation equation~\eqref{eq:mass_conservative} for the analytical flux (or cell values) $J_x\propto q n\varw_x$ can be shown to be equivalent to Faraday's law; in the numerical case this equivalency is shown using the discretized conservation equation and corresponding numerical flux $J_x\propto q F_n(\tilde{\vec{U}}) \simeq q n \varw_x$ for the current density $J_x$.
Taking the time derivative of equation~\eqref{eq:ExGauss} in conjunction with the discretized density update equation~\eqref{eq:FVUpdate} leads to the expression
\begin{align}
&\frac{\left(E_{x}\right)^{n+1}_i - \left(E_{x}\right)^n_i}{\Delta t}+\frac{ \left(E_{x}\right)^{n+1}_0- \left(E_{x}\right)^n_0}{\Delta t} = \nonumber\\
&\sum_s \frac{q_s}{\epsilon_0\Delta t}\int_{t_n}^{t_{n+1}}\left[-( F_{n, s})_{i} + (F_{n, s})_{0}\right] \mathrm{d} t.
\end{align}
 The integration in time using Runge-Kutta methods is the same as used to solve equation~\eqref{eq:FVUpdate}.
 Faraday's law using fluxes in one spatial dimension is then given by
\begin{equation}
	\label{eq:ExFaradayFlux}
	\left(E_{x}\right)^{n+1}_i = \left(E_{x}\right)^n_i - \sum_s \frac{q_s}{\epsilon_0} \int_{t_n}^{t_{n+1}} \left[ F_n\left(\tilde{\vec{U}}\right)\right]_{i, s} \mathrm{d} t,
\end{equation}
and enables us to identify $J_x$ by comparison to the charge conservation equation (equation~\ref{eq:mass_conservative} multiplied by $q_\mathrm{s}$)
\begin{equation}
	(J_{x})^{n+{1/2}}_{i} =\sum_s \frac{q_s}{\Delta t} \int_{t_n}^{t_{n+1}} \left[ F_n\left(\tilde{\vec{U}}\right)\right]_{i, s} \mathrm{d} t.
\end{equation}
Note, that the numerical flux also includes numerical diffusion and is directly related to changes in $\rho$. 
Due to this, other formulations for $J_x$ violate the charge conservation equation and can lead to numerical instabilities.

\subsubsection{Magnetic field evolution}
Because the fluid evolution influences the magnetic field only indirectly, the finite-difference time-domain (FDTD) update for the magnetic field is unchanged from the previous SHARP code.
For completeness we reproduce the formulae here \citep{shalabyNewCosmicRaydriven2020}
\begin{align}
	(B_y)_i^{n+\onehalf} &= (B_y)_i^{n-\onehalf} + \frac{\Delta t}{\Delta x}\left( (E_z)_{i+\onehalf}^{n} -  (E_z)_{i-\onehalf}^{n} \right),\\
	(B_z)_i^{n+\onehalf} &= (B_z)_i^{n-\onehalf} - \frac{\Delta t}{\Delta x}\left( (E_z)_{i+\onehalf}^{n} -  (E_y)_{i-\onehalf}^{n} \right).
\end{align}
$B_x$ is constant in the 1D3V model because of the requirement $\div\vec{B}=0$.

\subsection{Landau closure for fluid species}
\label{sec:heatflux}

The highest retained fluid moment, which is in our case the specific heat flux $\vec{Q}$, is not evolved in our set of equations. Instead, we need to estimate its value dynamically using an appropriate closure. 
The simple ideal gas closure sets $\vec{Q}=\mathbf{0}$, which, however, prevents the energy dissipation of plasma waves.
One important mechanism of such a dissipation is the collisionless damping of electrostatic waves achieved through Landau damping. 
Landau damping is a microphysical kinetic wave-particle interaction, where particles resonate with the wave exchange energy as a function of time. 
In essence, the resonant particles accelerate or decelerate to approach the wave's phase velocity, thereby picking up energy or releasing it, respectively.
For Maxwellian phase space distributions, there are more particles at velocities smaller than the phase velocity, which yields a net damping, i.e., energy loss of the wave \citep{boydPhysicsPlasmas2003}.

Various attempts, e.g.\ by \citet{hammettFluidMomentModels1990}, were carried out to approximate the heat flux $\vec{Q}$  of an almost Maxwellian distributed plasma, such that the kinetic phenomenon of Landau damping is mimicked in the linearized fluid equations.
Landau damping is a non-isotropic effect, which can be reflected in the fluid descriptions.
Accounting for the gyrotropy of the system around the magnetic field, often the double-adiabatic law with two adiabatic coefficients parallel and perpendicular to the magnetic field is presupposed \citep{hunanaIntroductoryGuideFluid2019}.
For now, we restrict our algorithm to isotropic pressures with only one common adiabatic coefficient for parallel and perpendicular pressure and leave this possibility of modelling anisotropic double-adiabatic systems open for future extensions of our algorithm.
In our simplified model, we denote an isotropized pressure tensor with the adiabatic coefficient $\Gamma = 5/3$, instantly isotropizing all heating occurring due to the heat flux closure, while $\Gamma = 3$ denotes a negligible pressure in the $y$ and $z$-direction.
Hence, we define only the perturbed scalar heat flux parallel to the magnetic field line $Q = Q_{\parallel}$ and no perpendicular heat flux.

Here, we will introduce two different formulae for heat flux closures.
The first and most popular collisionless electrostatic closure was proposed by \citet{hammettFluidMomentModels1990}. We refer to it as the $R_{32}$ closure throughout this paper, and it approximates the heat flux at a fixed $\Gamma = 3$, in Fourier space, by
\begin{equation}
\hat{Q} = - \ci\,\mathrm{sign}\left(k\right) \frac{2}{\sqrt{\upi }} \sqrt{2 \theta_0} c  n_0 \kB \frac{\hat{T}}{m}\equiv\hat{Q}_T.
\label{eq:R32}
\end{equation}

Here, hats are used to denote quantities in Fourier space along the magnetic field line, i.e.\ $\hat{Q}=\mathcal{F}_\parallel(Q)$, and the subscript $0$ refers to simulation box averages, that is $n_0= \sum_{i=0}^{N_{\rm c}} n_i/ N_{\rm c}$ is an average over all $N_\mathrm{c}$ cells.
Furthermore $\kB$ is the Boltzmann-constant, and  $\kB \hat{T} = \left( m \hat{p} - \kB T_0\hat{n}\right)/n_0$.
Since the plasma average or equilibrium temperature evolves slowly as a function of time, we adjust the background temperature $T_0$ after every time step to synchronize it with the mean pressure, $\kB T_0(t)/m = p_0 (t) / n_0$, while the density conservation ensures that $n_0$ stays constant.
Note also, that $Q_0=0$.
The dimensionless mass-normalized temperature is $\theta_0 = \kB T_0/(m c^2)$.

A more recent approximation was proposed by \citet{hunanaNewClosuresMore2018}, who restricts this closure to $\Gamma=3$ only, for reasons mentioned already.
We use an ad hoc formulation of their closure with a variable $\Gamma$, thereby allowing our simplified model to be used.
They also introduce the nomenclature $R_{mn}$ adopted here, which is used to denote that the kinetic plasma response function $R$ is mimicked for this closure by a Padé approximant with polynomials $P_m/Q_n=R_{mn}$ of order $m$ and $n$.
 We refer to their closure as $R_{31}$ and it approximates the heat flux, in Fourier space, by
\begin{equation}
\hat{Q} = \underbrace{\left(\frac{4}{4-\upi} - \Gamma\right) p_0 \hat{\varw}}_{\hat{Q}_\varw} +   \underbrace{\left(-\ci\,\mathrm{sign}\left(k\right) \frac{\sqrt{2\upi\theta_0}}{4-\upi} c n_0   \frac{\kB \hat{T}}{m}\right)}_{\hat{Q}_T}.
\label{eq:R31}
\end{equation}

In comparison to the $R_{32}$ closure, this closure has an additional dependence on the perturbed bulk velocity $\hat{\varw}$. 
This effectively increases the speed of sound obtained from the non-electromagnetic fluid equations and allows retrieving the correct damping rate with our ad-hoc assumption of variable $\Gamma$, see appendix~\ref{sec:adiabatic}.
For $\Gamma = 3$, we retrieve the coefficient for $\hat{\varw}$ from the aforementioned literature $\frac{4}{4-\upi} - 3 = \frac{3\upi - 8}{4-\upi}$.

In only one spatial dimension, as assumed in our code, the global integration along a magnetic field line is approximated to be along the spatial direction, i.e.\ $\mathcal{F}_\parallel=\mathcal{F}_x$.
An extension to multiple spatial dimensions with an anisotropic pressure tensor is not straightforward because in this case, this approach can lead to spurious instabilities \citep{passotFluidSimulationsIon2014} and the integration would need to be carried out along magnetic field lines.

A kinetic code does not need global communication to accurately reproduce Landau damping, since each particle (or particle bin) tracks its own interaction with each wave mode as a function of time and accumulates this information in the particle velocity.
However, after integrating out the individual particle velocities when building the evolution equations for the phase-space distribution function, i.e.\ equations~\eqref{eq:mass_conservative}-\eqref{eq:cons_energy},  information about the individual particle-wave interaction is no longer collected. Because some information about this interaction is also contained in the wave, such non-local information can be used to approximate the gradient of the physical heat flux, i.e., a closure of the fluid moments that incorporates such missing information.
This non-local information is approximated in equations~\eqref{eq:R32} and \eqref{eq:R31}, and is manifested by the term $\ci\, \mathrm{sign}\left(k\right)$ in Fourier space, which is also referred to as the
Hilbert transform.

Numerically, we do not include the heat flux in the Riemann solver used to compute the fluid fluxes. Instead, we compute the spatial derivative of the heat flux $\vnabla_\parallel \bcdot \vec Q$ separately.  We use Strang splitting for the $\vec \varw$ dependent part $\vec{Q}_\varw$ and the temperature dependent part $\vec{Q}_T$ to expand equation~\eqref{eq:splitting} into
\begin{equation}
	\label{eq:splitting_detailed}
\vec{U}^{n+\onehalf} = \mathrm{e}^{\frac{\Delta t}{2} \mat{F}} \mathrm{e}^{\frac{\Delta t}{2} \vec{Q}_\varw} \mathrm{e}^{\Delta t \vec{Q}_T}   \mathrm{e}^{\Delta t \vec{S}_{\rm em} } \mathrm{e}^{\frac{\Delta t}{2} \vec{Q}_\varw} \mathrm{e}^{\frac{\Delta t}{2} \mat{F}} \vec{U}^{n-\onehalf} + O(\Delta t^3),
\end{equation}
such that only one non-global evaluation of $\vec{Q}_T$ is needed.
Using Heun's method together with the fast Fourier transform (FFT) the update formulae for the pressure w.r.t.\ operators $\vec{Q}_\varw$ and $\vec{Q}_T$ are respectively
\begin{align}
  \left.p^{n+1}\right|_{Q_\varw} &= \mathrm{e}^{\Delta t Q_\varw} p^n
  = p^n +  \Delta t a_\varw p_0 \vnabla_\parallel \bcdot \vec\varw, \\
  \label{eq:pressureUpdateTemp}
  \left.p^{n+1}\right|_{Q_T} &= \mathrm{e}^{\Delta t Q_T} p^n
  = p^n + \Delta t \mathcal{F}^{-1}_\parallel
  \left[\abs{k} a_T \left(1+\frac{\Delta t}{2} \abs{k} a_T\right) \hat{T}^n  \right] ,
\end{align}
where the derivative in Fourier space was obtained by multiplying with $\ci k$ and the inverse FFT is denoted by $\mathcal{F}^{-1}$. 
For the $R_{31}$ closure the coefficients are given by $a_\varw=4/(4-\upi)$ and $a_T= (4-\upi)^{-1} (2\upi \theta_0)^{1/2} c n_0 \kB/m$, while for the $R_{32}$ closure these are given by $a_\varw = 0$ and $a_T=2 (2 \theta_0/\upi )^{1/2} c  n_0 \kB /m$.
Both closures compute a term proportional to $\hat{T}$ (cf.\ equation~\ref{eq:pressureUpdateTemp})
\begin{equation}
	\label{eq:heatfluxDerProp}
	\ci k \hat{Q} \propto - \ci\, \mathrm{sign}\left(k\right)  \ci k a_T \hat{T} =   \abs{k} a_T \hat{T}. 
\end{equation}
Computing this term naively using the FFT is expensive.
This is why, in the following, we present local, semi-local, and efficient global (Fourier transform-based) numerical approximations of the Landau closures, which we have implemented in the fluid-SHARP code.

\subsubsection{Local approximations of the Hilbert transform}
The phase shift between the wanted derivative $\ci k \hat{Q}$ and the input of $\hat{T}$ in equation~\eqref{eq:heatfluxDerProp} is exactly $0$,
while the amplitude is proportional to $\abs{k}$.  This is therefore a special case ($a=1$) of the fractional Riesz derivative
$\partial^{a}/\partial\abs{x}^a$ with Fourier representation
\begin{equation}
\mathcal{F}\left( \pdv[a]{f\left(x\right)}{\abs{x}} \right) = -\abs{k}^a \hat{f}\left(k\right),
\end{equation}
where $a \in \mathbb{R}$. 
Note, that all approximations mentioned here only introduce errors in the
amplitude of $\abs{k}$, but not in its phase.  This makes them easier to integrate into
simulations in comparison to approximations which are not designed to prevent
phase errors, because large phase errors  (between
$\uppi/2$ and $3\uppi/2$) in any wave mode transform the damping term into an exponentially growing numerical instability.
The local approximations make use of the fact, that the fractional Riesz derivative is local and cheap to evaluate for the special case
$a = 2m$ with $m \in \mathbb{N}^0$, where it reproduces the usual
derivative $\partial^{2m}/\partial{\abs{x}}^{2m} =
\left(-1\right)^{m+1}\partial^{2m}/\partial x^{2m}$.
\citet{wangComparisonMultifluidMoment2015} use $a=0$, while
\citet{allmann-rahnTemperatureGradientDriven2018} and
\citet{ngImprovedTenmomentClosure2020} approximate the non-isotropic pressure tensors with $a=2$. These approximations are scaled to a characteristic wavenumber $k_0$ at which the damping is expected to occur. 

\begin{figure}
	\centering
	\includegraphics{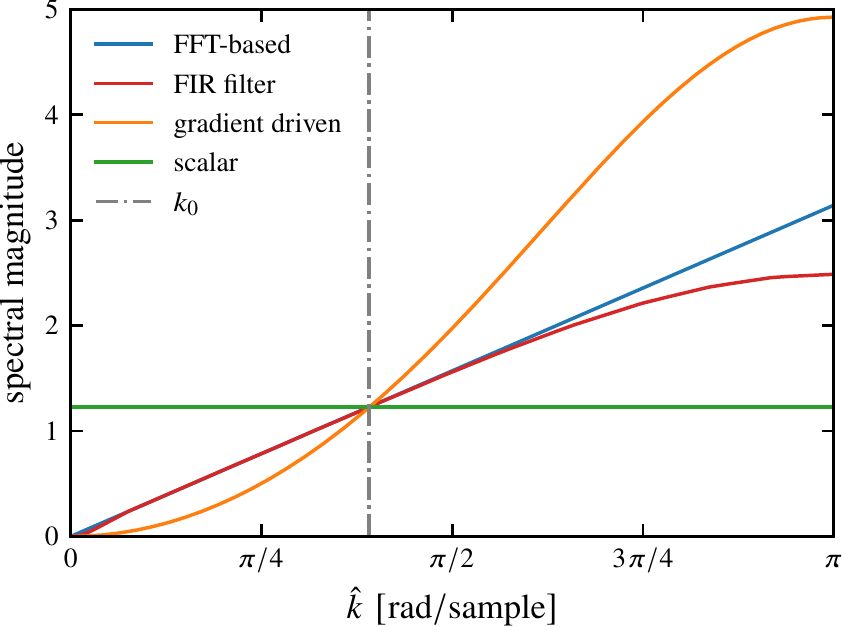}
	\caption{The magnitude of the frequency response, which is a quantification of how much the amplitude at a specific frequency is amplified or suppressed, of different approximations of the derivative of the Hilbert transform. $\hat{k}$ is given in normalized frequencies (with regards to the Nyquist frequency), while the negative frequencies in the interval $\left[-\uppi, 0\right]$ are not shown here due to the symmetric dependence of all plotted values on $\abs{k}$.
		The FFT-based approach reproduces the correct, linear response.
	The scalar and gradient driven closures are given by equations~\eqref{eq:heatfluxApproxScalar} and ~\eqref{eq:heatfluxApproxQuadratic} respectively with the parameter $k_0$ marked as a grey, vertical line.  
	The FIR filter is described by equation~\eqref{eq:filterHilbert}.
	}
	\label{fig:heatfluxApproxs}
\end{figure}

 The choice of $a=0$ means, that the
approximation is a scalar
\begin{equation}
	\label{eq:heatfluxApproxScalar}
\ci k \hat{Q}  \propto \abs{k_0} \hat{T}, 
\end{equation}
while the gradient-driven closures with $a=2$ use
\begin{equation}
	\label{eq:heatfluxApproxQuadratic}
\ci k \hat{Q} \propto \frac{k^2}{\abs{k_0}} \hat{T} .
\end{equation}
The gradient-driven closures are equal to the FFT solution at two wavelengths,
$0$ and $k_0$, while the scalar closure is only exact at $k_0$, see
Fig.~\ref{fig:heatfluxApproxs}.  
Since $\ci k\hat{Q}$ is not computed alongside with the conservative fluxes in the Riemann solver, energy conservation is only preserved if the mean energy does
not increase. To achieve this, the approximation for the derivative of the heat flux needs to vanish at wavenumber $0$, which the scalar approximation does not fulfil.

Because fluid closures are only approximately mimicking kinetic Landau damping anyway,
these local approximations to the fluid closures are useful to save computational cost.
Furthermore they are easier to implement, especially when the full pressure tensor is computed.
 However, they may lead to misleading results in multiscale simulations, where multiple characteristic damping lengths are present and depend on the estimate of $k_0$.
For example, \citet{allmann-rahnEnergyConservingVlasov2021} show a case where ion and
electron heating intensities are switched qualitatively.

\subsubsection{Semi-local approximations of the Hilbert transform}
While the less accurate local approximations use an arbitrary value of $k_0$, the FFT is expensive and depends on periodic boundary conditions.
Here, we aim to have a fallback algorithm as a compromise between both approaches.

A digital finite impulse response (FIR) filter can be designed to approximate the non-local effects
by convolving the simulation data with adjacent auxiliary data points, where the filter length determines the maximum distance.
For example, an asymmetric filter with an even number of entries is applied on an input $x$ using filter coefficients $b_j$, producing the output $y$:
\begin{equation}
	\label{eq:filter}
	y_{i+0.5} = \sum_{j=-(N_f/2-0.5)}^{N_f/2-0.5} b_j x_{i+j+0.5} .
\end{equation}
A numerical derivative is then an asymmetrical filter with $N_f=2$ and coefficients $b_{\pm0.5} = \pm /\Delta x$, such that $y_{i+0.5} = (x_{i+1} - x_i)/\Delta x$.
Figure \ref{fig:heatfluxApproxs} shows the magnitude of the frequency response.
The gradient driven case shows a quadratic $k^2$ dependence, which is suppressed for larger $k$.
This is due to the relatively small uneven filter length of $7$ used here;
the filter length is an important parameter, since it influences the accuracy of the approximation.
With a filter length corresponding to the simulation box size the results can converge to the FFT-based algorithm (i.e.\ the $k^2$ dependence is not suppressed at higher $k$), if the filter is designed appropriately.
As noted previously, the local closures do not converge to $\partial/\partial\abs{x}$.
A correct convergence for approximating $\partial/\partial\abs{x}$ is obtained through the high order formulation by \citet{dingHighorderAlgorithmsRiesz2015}.
However, this filter violates energy conservation for smaller filter length and is thus, not suitable for our case.
Instead, we construct the filter by adopting a convolution of two sub-filters, each of which has an odd amount of asymmetric entries (termed a Type IV filter) similar to the numerical derivative mentioned already.
By design, their output has a vanishing mean, thereby guaranteeing energy conservation.
A symmetric splitting into the sub-filters $\partial/\partial\abs{x}  = (\partial^{1/2}/\partial\abs{x}^{1/2} )^2$ is possible, however its frequency response is not monotonic (and has visible ripples) for small filter lengths. This leads to the nonphysical case that some waves at a particular wavenumber $k$ are damped less than their slightly larger scale waves at $k-\delta k$.

Instead, we opt to use the intuitive splitting of $\partial/\partial\abs{x}=\partial/\partial{x} \mathcal{H}$ where the Hilbert-transform filter $\mathcal{H}$ is equivalent to $-\ci\, \mathrm{sign}\left(k\right)$ in Fourier space.
The filter $\mathcal{H}$ has coefficients $b_{j}=1/(\upi j)$. 
We derive an equivalent formulation to equation~\eqref{eq:pressureUpdateTemp}, which is first order in time, by applying the derivative and Hilbert-transform filters successively, i.e.\
\begin{equation}
	\label{eq:filterHilbert}
	p^{n+1} = p^n + \Delta t a_T \pdv{}{x} \sum_{j=-(N_f/2-0.5)}^{N_f/2-0.5} \frac{1}{\upi j} T^n_{i+j+0.5}.
\end{equation}
Note, that the derivative is also computed by convolution and has a separate filter length corresponding to its spatial order. We opt to use the same spatial order as in the C-WENO reconstruction for the finite volume scheme.

Even for small Hilbert-transform filter lengths in comparison to the number of cells, e.g.\ $N_f/N_{\rm c} = 0.04$ as shown in figure~\ref{fig:heatfluxApproxs}, this formulation dramatically improves the accuracy of multiscale problems in comparison to local approximations.
Here, $N_f$ is critical for the accuracy at small wavenumbers $k$, while the spatial order of the derivative is critical for the accuracy at large $k$.
Most importantly, this semi-local approach does not require setting an arbitrary damping scale $k_0$ such as the local approximations mentioned before. The only parameter of this approach is the filter length, which should be chosen to be sufficiently large.

\subsubsection{Efficient FFT-based computation of the Hilbert transform}
\label{sec:fftpar}
Provided the plasma background is uniform and periodic, the most accurate while
computationally most expensive results are achieved by computing the heat flux
of the fluid in Fourier space.  
While the FFT is easy to compute on a single computer using standard numerical libraries, our code is parallelized using MPI and an efficient one-dimensional FFT is needed. 
The computation of the Fourier transform is
expensive for two reasons:
\begin{enumerate}
	\item globally, each Fourier component needs to be informed about data from
          every other computational cell (which may be stored on a different
          processor), and
	\item the Fourier transform is not easily parallelizable in one
          dimension, which precludes an efficient scalable Fourier algorithm.
\end{enumerate}
This naturally limits the overall computational scalability of the fluid part of
the code. Communication over multiple MPI processes is time consuming because of
latency and finite bandwidth.
For this reason, parallel FFT algorithms are prone to become a computational bottleneck. However, using non-blocking MPI routines to perform communication in the background can be used while the high computational load of the particles is carried out.
Thus, in our case of a combined fluid and PIC algorithm, the communication required for an accurate FFT-based heat flux computation is comparatively computationally cheaper,  even with relatively small numbers of PIC particles.
Hence, in our case the FFT algorithm does not necessarily become a bottleneck for larger problems.

In order to distribute the computational load of the FFT, we employ a
four-step algorithm in the first step of the computation
\citep{baileyFFTsExternalHierarchical1990,
 takahashiHighPerformanceRadix2Parallel2000}, which extends the Cooley-Tukey algorithm \citep{cooleyAlgorithmMachineCalculation1965} for multiple processors. We shortly describe the algorithm for complex input data as found in the literature and afterwards adapt the parallel FFT for real input data in our implementation.
 The four-step algorithm interprets the complex data vector $x_j$ of length $N$ as a two-dimensional vector $x_j= x_{j_1, j_2}$ with lengths $n_1$ and $n_2$ respectively, and volume $n_1 n_2 = N$.
The mapping $j = j_1 + j_2n_1$ and $k = k_2 + k_1n_2$ is inserted into the
definition of the discrete Fourier transform, where $\Psi = \exp{-2 \upi \ci}$
\begin{align}
\hat{x}_k &= \sum_{j=0}^{N-1} x_j \Psi^{jk/N}, \\
\hat{x}_{k_2,k_1}&= \sum_{j_1 = 0}^{n_1-1} \sum_{j_2 = 0}^{n_2-1} x_{j_1,j2} \Psi^{ j_2k_2/n_2} \Psi^{j_1k_2/N} \Psi^{j_1k_1/n_1}  .
\end{align}
This way, a complex-to-complex parallel FFT of length $N$ is distributed to $n_1$ local FFTs of length $n_2$, a
multiplication by the twiddle factors $\Psi^{j_1k_2/N}$ and finally $n_2$ FFTs of length $n_1$, with a communication intensive transpose in between.
All-to-all communication takes place two times, in the first step -- cyclically distributing $j$ to $j_1$ and $j_2$ -- and for the transpose.
A third all-to-all communication would be needed to properly sort the values in Fourier space. However, a scrambled output suffices for computing the heat flux.
Furthermore, since often two FFTs, i.e.\ electrons and ions, need to be computed simultaneously, they can be computed on different nodes. This has the advantage, that the second all-to-all communication for the transpose is not completely global resulting in reduced communication times.

Adapting this algorithm to a real-to-complex FFT, where due to Hermitian symmetry only values of $k\leq\lfloor N/2\rfloor$ need to be computed, a large amount of computational and communicational savings can be realized.
A real-to-complex parallel FFT of length $N$ is distributed to $n_1$ local real-to-complex FFTs of length $n_2$, a
multiplication by the twiddle factors $\Psi^{j_1k_2/N}$ and, now only, $\lfloor n_2/2\rfloor + 1$ complex-to-complex FFTs of length $n_1$.
Up to two of the latter FFTs can be replaced by real-to-complex FFTs, along the axes $k_2=0$ and, if $n_2$ is even, $k_2=n_2/2$.
A scrambled output is received, which, due to Hermitian symmetry, needs to be partially complex conjugated.

A key point in ensuring the efficiency of the parallel four-step algorithm consists in choosing large $n_1$ and $n_2$.
$n_1 \simeq n_2 \simeq \sqrt{N}$ is the optimal choice for the distributed complex-to-complex FFT, the real-to-complex FFT should prefer $n_1 \simeq \lfloor n_2/2\rfloor +1\simeq (\sqrt{2N+1} + 1)/2$.
The computational scaling with $P$ processors and roughly optimally distributed $n_1$ and $n_2$ is akin to $\textit{O}\left(N/P\log{N}\right)$, but degrades if $N$ is a prime number, or, more generally, if $n_1$ or $n_2/2$ is smaller than the number of processors. This easily avoidable because $N$ is a free parameter, and so are $n_1$ and $n_2$.
While this does not scale favourably in comparison to the $\textit{O}\left(N/P\right)$ scaling that dominates the rest of the fluid code, still, the FFT is trivially independent
of the numbers of particles per cell $N_\mathrm{pc}$. The PIC-module on the other hand scales as $\textit{O}\left(N_\mathrm{pc} N/P\right)$ and typical applications have $N_\mathrm{pc}\gtrsim100$.
In many applications the cost of the Fourier transform is, even with worse scaling, subdominant in comparison to the cost of the PIC part.
In the remaining cases, local approximations, discussed above, are favourable.

\subsection{Current-coupled fluid-PIC algorithm}
\label{sec:fluid-PIC} 
The coupling in our code between various fluid and kinetic (PIC) species is achieved through a current-coupling scheme.
Namely, both fluid and kinetic species contribute to the charge and current densities.
The electromagnetic fields then evolve in response to the total contributions.
The fields are staggered on a Yee-mesh and are updated with the FDTD scheme. 
Subsequently, both fluid and kinetic species evolve in repose to the new electromagnetic fields. That is our current-coupling scheme does not make any assumption on the velocity distribution of the species modelled using the kinetic description~\citep{Park1992}.

The PIC species, using fifth-order spline interpolation, are deposited to specific points on the Yee-grid for which the charge density is defined at full-time steps while the current density is defined at half-time steps as discussed by~\citet{shalabySHARPSpatiallyHigherorder2017, shalabyNewCosmicRaydriven2020}.
For fluid species, the fluid density and velocity are defined at the same time step. Therefore, during the evolution of the fluid, we deposit the fluid contribution to the charge and current densities, $\rho$ and $\vec{J}_{y, z}$ respectively, at the cell centres.
The deposition for $\vec{J}_{y, z}$ is trivial at half-time steps, where the fluid vector $\vec{U}$ is defined, while the contribution to $\rho$ is computed at full-time steps, i.e.\ before the electromagnetic source update according to equation~\eqref{eq:splitting}.
Note, that $\rho$ stays constant when computing the Lorentz force and heat flux updates.

Our algorithm does not apply any approximations to the electrical field components or to Ohm's law, requiring electron timescales and motions to be fully resolved. 
Consequently, we apply the same algorithm to fluid electrons and protons. This is accomplished using the modular design of the fluid SHARP code where each fluid species is represented by initialising a fluid code class. Each instance of this code class is initialized using the values of the mass and the charges of their respective particle species. The algorithms which define the evolution of each particle species are implemented as functions of the fluid class. This allows us to setup simulations with multiple species, all of which are evolved with the same numerical algorithms, with little effort.

\begin{figure}
	\centering
		\includegraphics[width=0.8\textwidth]{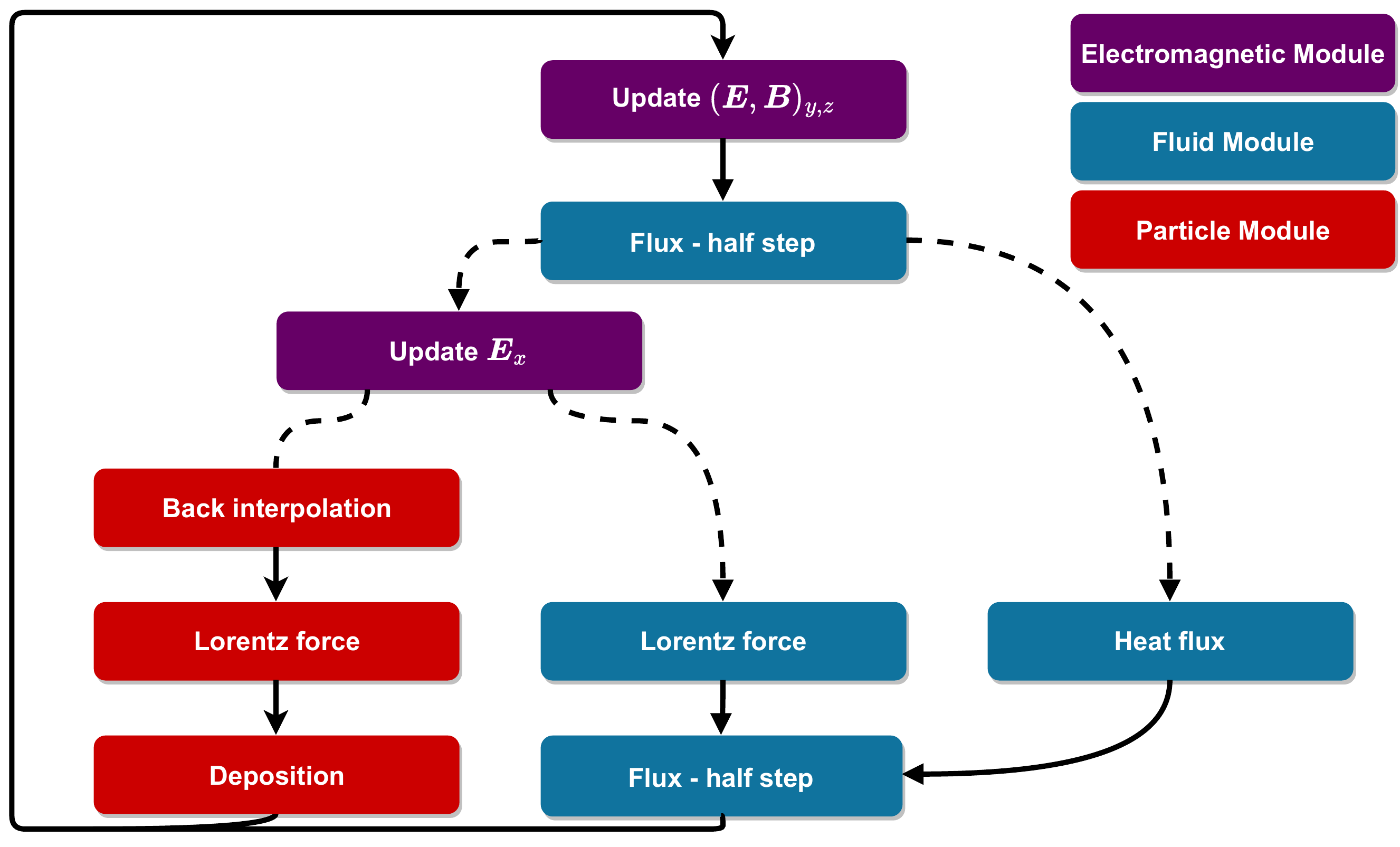}
	\caption{Schematic representation of the interaction of the different
          modules in the fluid-SHARP code.  Red boxes belong to the particle
          class, violet boxes to the electromagnetic class and blue boxes to
          the fluid class.  Dashed lines show branches which are
          task parallelizable, i.e.\ where non-blocking MPI communication can be used for overlapping communication and computation. }
	\label{fig:Code}
\end{figure}

In figure~\ref{fig:Code} the main loop of the fluid-PIC algorithm is presented.
It can be seen that the usual PIC-algorithm loop of electromagnetic update, interpolation to particle position, particle push, and field deposition is retrieved when no fluid species is initialised.
On the other hand, without PIC particles, we retrieve a multispecies fluid plasma code.
While our fluid-PIC algorithm can simulate an arbitrary mixture of species, it is most efficient if fluids are used for background species and particles for non-thermal particle distributions.  
Possibilities for task parallelization are shown in figure~\ref{fig:Code} by dashed lines, which allows maximizing computation-communication overlap.

Our fluid implementation is included within the SHARP code, which uses a fifth-order spline function for deposition and back-interpolation for PIC species~\citep{shalabySHARPSpatiallyHigherorder2017, shalabyNewCosmicRaydriven2020}. 
The PIC part of the code does not make use of filtering grid quantities and results in comparatively small numerical heating per time step, which (if present) would affect the reliability of the simulation results on long timescales (see section~5 in \citet{shalabySHARPSpatiallyHigherorder2017}).
This property is important because we are specifically interested in studying microphysical effects on long timescales with our fluid-PIC code.
Due to the modularity of our code, each part can be tested individually. These tests, ranging from the uncharged fluid solver to full fluid-PIC simulations, are shown in the next section.

\section{Code validation tests}
\label{sec:results}
In this section, we present the results of various code tests. We start with two
shock-tube tests in section~\ref{sec:shock_tube} before we show that our code is able to accurately capture all six branches of the two-fluid
dispersion relation (Section~\ref{sec:2fluidDR}). We describe code tests of
Langmuir wave damping (Section~\ref{sec:Langmuir_damping}) and of two
interacting Alfv\'en waves generating a new, longitudinal wave along the
magnetic field (Section~\ref{sec:Alfven_waves}). In section~\ref{sec:CSRI}, we
test the entire fluid-PIC code with a simulation of the gyrotropic CR
streaming instability, where PIC CRs are streaming in a stationary
electron-proton fluid background. Finally, we demonstrate the successful
parallelization strategy of our code by performing scaling tests in
Section~\ref{sec:scaling}.

\subsection{Shock tube}
\label{sec:shock_tube}
\begin{figure*}
\subfloat[	\centering Shock tube test 1, a modified Sod shock tube, at time 0.2 (code units).]{
	\includegraphics[width=0.5\textwidth]{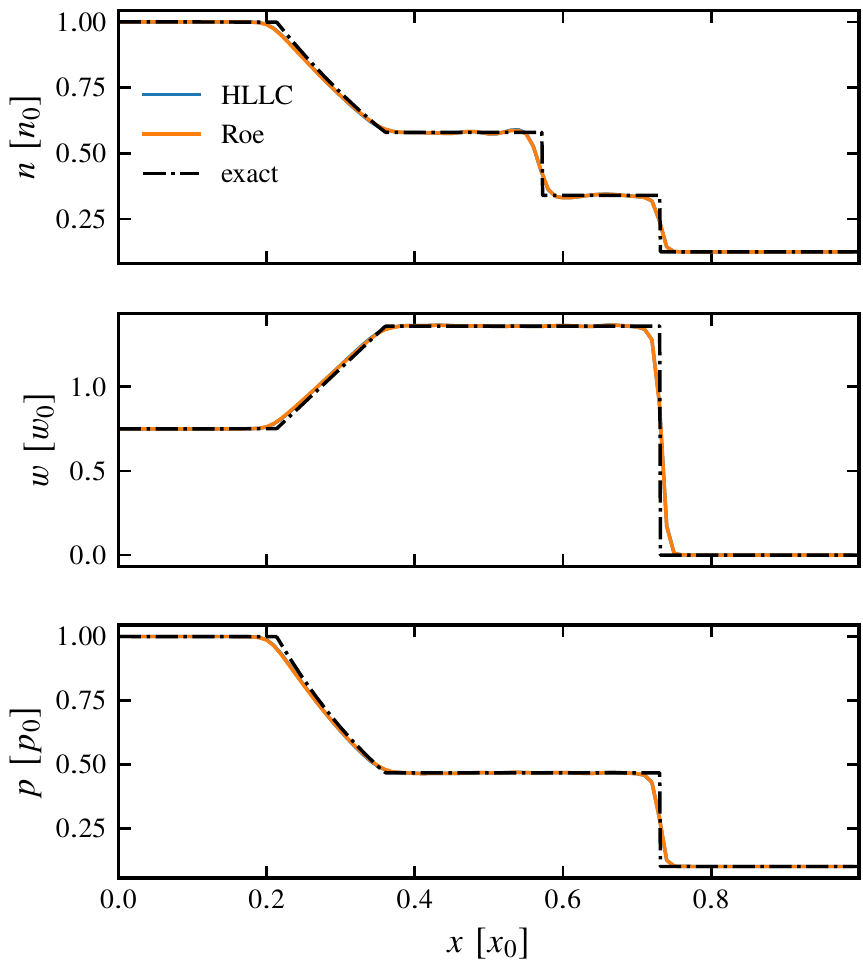}
	\label{fig:testFluid1}
}
\subfloat[Shock tube test 2 at time 0.012 (code units).]{
	\includegraphics[width=0.5\textwidth]{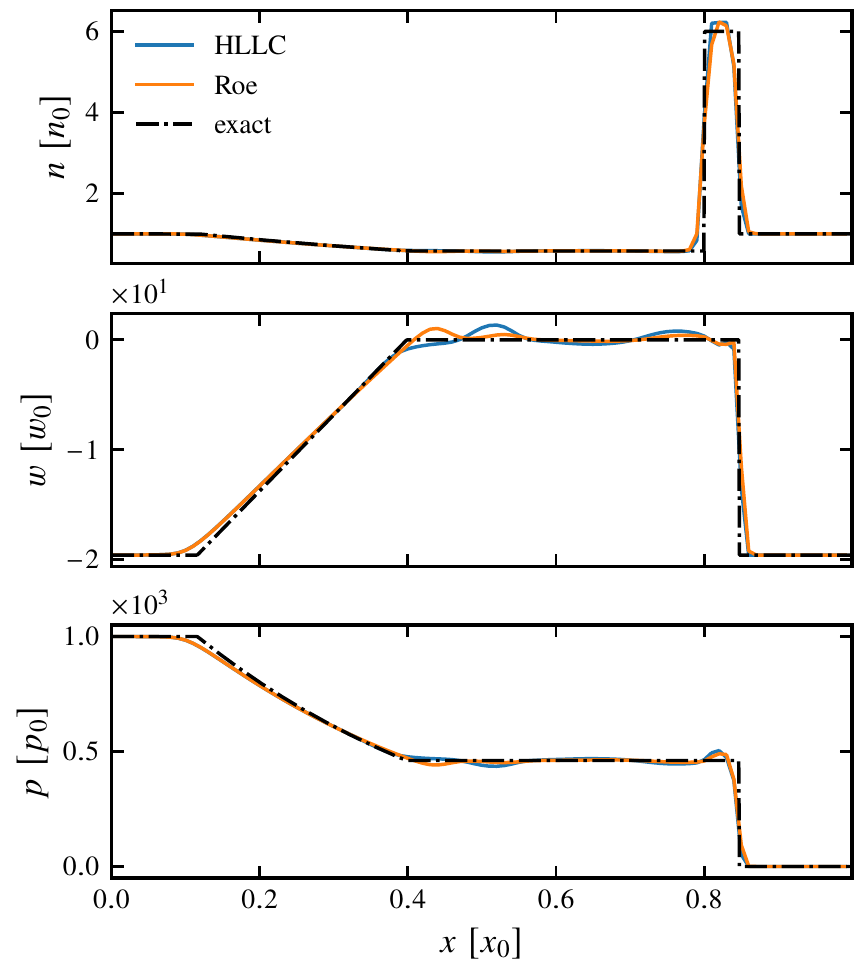}
	\label{fig:testFluid5}
}
\caption{1D1V hydrodynamical shock tube tests with initial conditions given in
	table~\ref{tab:shocktest}. The simulations carried out with the HLLC and Roe Riemann solvers are compared to the exact solutions. Density, bulk velocity in $x$-direction and pressure are plotted for each test.}
\end{figure*}
\begin{table}
  \begin{center}
  \begin{tabular}{c|c|*{3}{c}|*{3}{c}}
    Test & $x_0$ & $n_\mathrm{l}$ & $\varw_\mathrm{l}$ &  $p_\mathrm{l}$ & $n_\mathrm{r}$ & $\varw_\mathrm{r}$ &  $p_\mathrm{r}$ \\
    \hline
    \hline
    1 & 0.3 & 1 & 0.75 & 1 & 0.125& 0 & 0.1 \\
    2 & 0.8 & 1 & -19.59745 & 1000 & 1 & -19.59745& 0.01 \\
  \end{tabular}

  \caption{Parameters adopted for the shock tube tests described in section~\ref{sec:shock_tube}. $x_0$ divides the domain into two halves, where values to the left of $x_0$ ($x<x_0$) are initialized by the parameters with subscript $\mathrm{l}$. Similarly, subscript $\mathrm{r}$ indicates parameters to the right of $x_0$.}
    \label{tab:shocktest}
    \end{center}
\end{table}

As the fluid approximation will be primarily used for background plasmas without
excessive gradients, the accuracy of resolving sharp discontinuities is of
secondary importance in practical applications.  Still, we stress test our
implementation of the fluid equations to ensure its numerical robustness and to
compare the numerical dispersion for different Riemann solvers.  For the shock
tests a numerical grid of $100$ cells is used with a constant CFL number
$C_{\mathrm{cfl}}=0.2$ with the adiabatic coefficient $\Gamma = 1.4$.  The boundary
conditions are transmissive and the initial conditions for the tests are given
in table~\ref{tab:shocktest}, which are the same as in
\citet{toroRiemannSolversNumerical2009}, where a CFL number of $0.2\times0.95$
is used only in the first five steps and $0.95$ afterward.  The units used for
these non-electromagnetic tests are arbitrary units and do not coincide with the
usual simulation units.

Test 1, as shown in figure~\ref{fig:testFluid1}, is a modified Sod shock tube
test.  The sonic rarefaction wave on the left-hand side as well as the shock
front on the right are well resolved without noticeable oscillations. The
contact discontinuity in the middle introduces small oscillations in the density
and is smeared out more than the shock front.  While the Roe and HLLC solvers
yield almost the same results, the HLLC solver is slightly better at resolving
the sonic point at the head (to the left) of the sonic rarefaction wave, which
the Roe solver can only resolve because an entropy fix is applied.

Figure~\ref{fig:testFluid5} shows a test of a stationary contact discontinuity
with a shock front of a high Mach number travelling to the right and a rarefaction
wave to the left.  It can be seen, that while the HLLC method introduces more
oscillations, it is also better at resolving the contact discontinuity.

In low-density flows the Roe solver is not suitable because it is not robust
without further modifications \citep{einfeldtGodunovtypeMethodsLow1991}, making
the HLLC method slightly more robust while the Roe method is slightly less
dispersive.  However, for most practical applications studied here, both methods produce similar results.

\subsection{Two-fluid dispersion relation}
\label{sec:2fluidDR}
\begin{figure}
  \centering
  \includegraphics{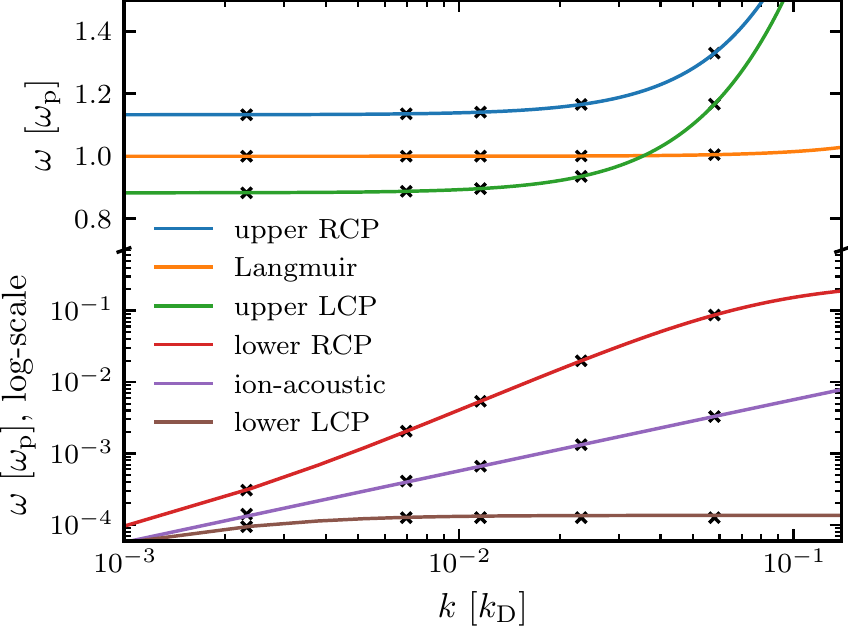}
  \caption{The six branches of the two-fluid dispersion relation are shown, with two electrostatic wave branches (Langmuir and ion-acoustic) as well as four electromagnetic left and right-hand circularly polarized wave branches (LCP and RCP). Often, the lower RCP is referred to as whistler branch and the lower LCP as ion cyclotron branch; for parallel propagation their phase velocities approach the Alfv\'en speed at small $k$.
  	The upper RCP and LCP are modified light waves.
  	 We mark the six local extrema of the Fourier-transformed fluid simulation outputs at each wavenumber with crosses. Theoretical predictions are shown as lines.  }
    \label{fig:testDisp}
\end{figure}

\begin{figure}
  \centering
   \includegraphics[width=\columnwidth]{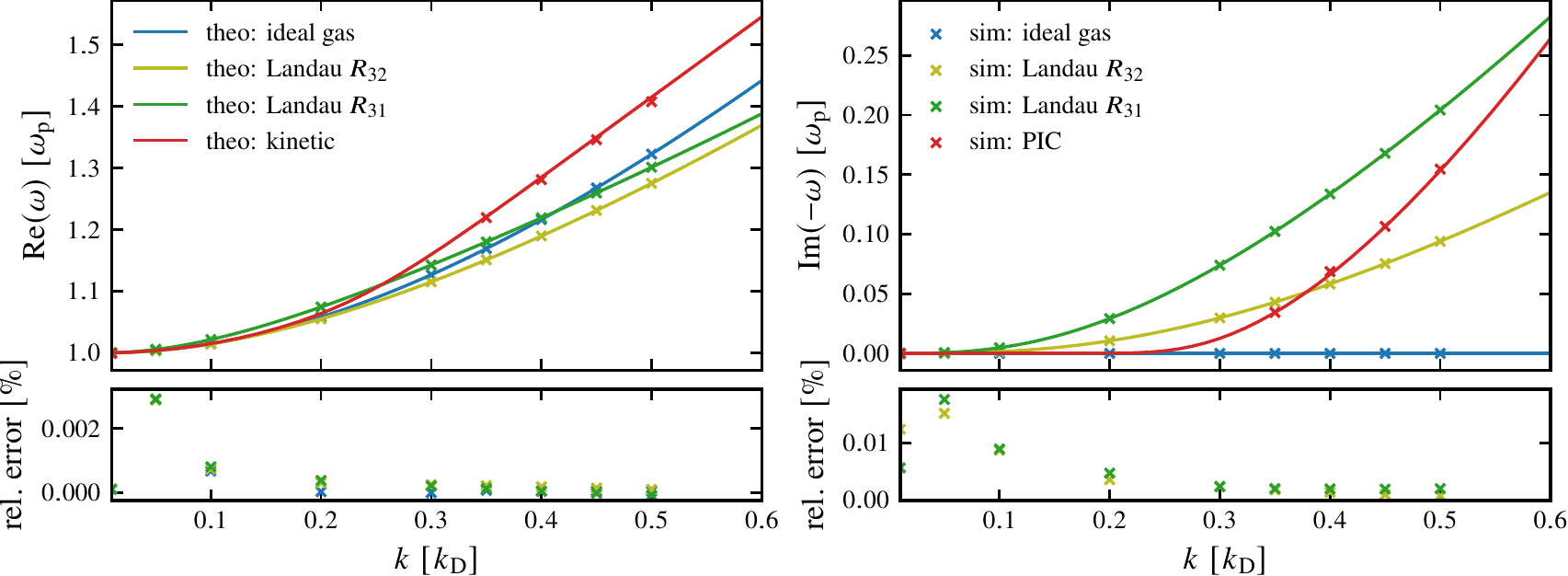}
  \caption{The linear dispersion relations of a Langmuir wave with immobile
    ions. Shown are, on the left-hand side, the real frequency components and,
    on the right-hand side, the negative imaginary frequency components (which
    are responsible for damping).  The crosses present data points obtained from
    simulations with the respective closure while the theoretical result is
    shown with a solid line. The relative error between simulation and
    theoretical results $(\omega^{\mathrm{sim}} - \omega^{\mathrm{theor}}) /
    \omega^{\mathrm{theor}}$ is shown in the lower panels.  For reference, the red
    crosses display the data points as given in table~1 of
    \citet{shalabySHARPSpatiallyHigherorder2017}.  }
  \label{fig:dp1f}
\end{figure}

For an ideal two-fluid plasma the dispersion relation can be solved for six
different wave branches~\citep{stixWavesPlasmas1992}.  We show the solutions to
the dispersion relation of a two-fluid plasma in figure~\ref{fig:testDisp} for a
realistic mass ratio of $m_\mathrm{i} = 1836 m_\mathrm{e}$ and $\beta_{\rm i} = n \kB
T_{\rm i}/[B_0^2/(2\mu_0)] = 0.2$ in an isothermal plasma. $B_0$ is oriented
along the $x$-axis and the Alfv\'en velocity is $\varv_{\rm A} = B_0/(\mu_0
n_\mathrm{i} m_\mathrm{i})^{1/2}=5.83\times10^{-3} c$.  Multiple simulations at
different wavenumbers have been initialized that have all six wave modes
simultaneously present and were run for a total time of
$14/\min\left(\omega\right)$, where $\omega$ denotes the wave frequencies, which
are always completely real for an ideal fluid. Consequently, the waves should be
undamped and any possible damping introduced is because of numerical
dissipation. Initial conditions for all of our fluid simulations as well as
theoretical predictions are computed using an extended algorithm based on the
dispersion solver by \citet{xiePDRFGeneralDispersion2014}, which can take into
account the effects of both heat flux closures.  A Fourier analysis in time has
been performed and the six largest local extrema are shown as crosses in
Fig.~\ref{fig:testDisp}.  It can be seen, that the simulation results are in
good agreement with the analytical results.  In the Fourier-analysis the largest
relative errors of at most $7$ per cent in $\omega$ occur in the large-scale
part of the ion-acoustic branch as well as close to the cut-off frequency of the
lower LCP branch. In comparison to this, the largest relative errors in the upper
three branches are more than one magnitude less.

\subsection{Langmuir wave damping}
\label{sec:Langmuir_damping}

The electrostatic wave modes are directly subject to linear Landau damping, and thus present a good test for the heat flux closures.
To test this, we initialize standing Langmuir waves in an electron plasma with immobile ions.
We use the same grid layouts as in
table~1 of \citet{shalabySHARPSpatiallyHigherorder2017}, supplemented with fluid simulations run at $k/k_\mathrm{D} \in \left\{0.1, 0.2, 0.3\right\}$ with a resolution of $\uplambda/\Delta x = 68$ cells per wavelength and a domain size of length $L = 10\uplambda$ wavelengths.
The wavenumber associated with the Debye length is the ratio of plasma frequency to thermal velocity, i.e.\ $k_\mathrm{D}=\omega_\mathrm{p}/ \theta^{1/2}c$.
The amplitude of the wave is chosen, such that the density fluctuation to background ratio is fixed to $\delta n/n_0=10^{-3}$.

In order to find the numerical dispersion relation we perform curve fitting with the Powell algorithm on the time series for times up to $80\,\omega_\mathrm{p}^{-1}$, while the simulations at $k/k_{\rm D}=0.01$ and $0.05$ with small damping are analysed up to $240\,\omega_\mathrm{p}^{-1}$. The computation of the heat fluxes for the $R_{31}$ and $R_{32}$ closures is performed using the FFT-based method.
The results are shown in figure~\ref{fig:dp1f}, where the ideal gas closure and the kinetic results are also depicted for reference.

Generally, it can be seen, that at small scales the closures show larger deviations from each other, which is also where the fluid description starts breaking down naturally as the particle distribution is not in equilibrium.  At larger scales, the various descriptions of Landau damping converge and approach zero.
The numerical relative error of the fluid code is small and stays below $0.003$ per cent for real frequencies and below $0.02$ per cent for decay rates in this setup.
The simulation at $k/k_{\rm D}=0.05$ performs worse than the one at $k/k_{\rm D}=0.1$ due to the significantly lower resolution. The error in $\omega$ decreases at second-order with increasing spatial resolution, as shown in appendix~\ref{sec:convergence}.

\subsection{Interacting Alfv\'en waves}
\label{sec:Alfven_waves}

\begin{figure}
  \centering
  \includegraphics[width=0.5\textwidth]{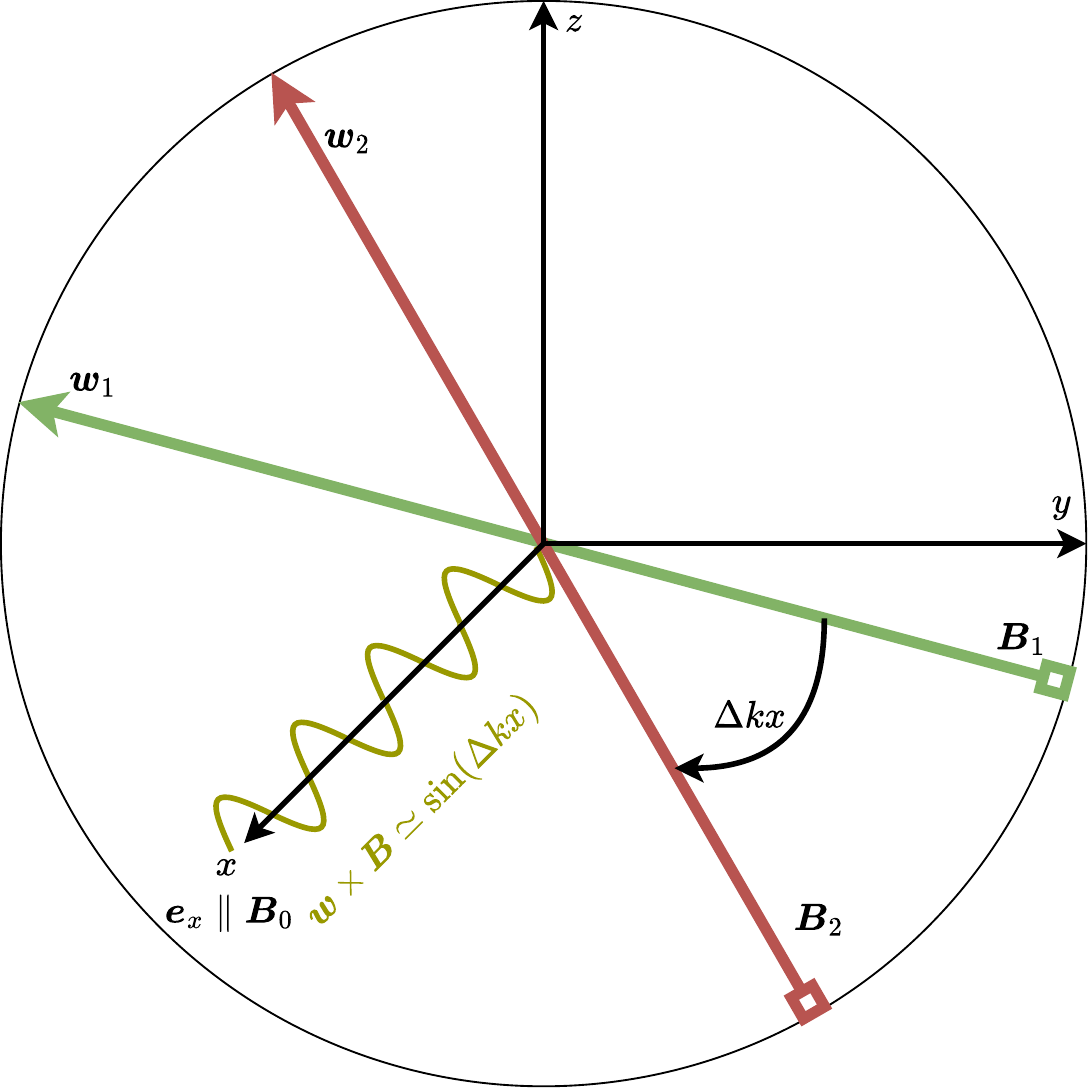}
  \caption{Two different Alfv\'en waves, with magnetic and velocity vectors
    $\vec{B}_1, \vec{B}_2$ and $\vec{\varw}_1, \vec{\varw}_2$, propagate
    transversally along the $x$-axis, where the electromagnetic vectors rotate
    (counter-)clockwise around it. Because of their phase difference $\Delta kx$
    the overall Lorentz force $(\vec \varw_1 + \vec \varw_2)\times(\vec B_1 +
    \vec B_2)$ in $x$-direction is non-zero, thereby generating the longitudinal
    wave shown in dark yellow.  }
  \label{fig:AlfInt}
\end{figure}

A single Alfv\'en wave is purely transversal and not directly affected by Landau
damping.  However, two or more Alfv\'en waves drive a longitudinal electrostatic wave, which
is susceptible to Landau damping, see figure~\ref{fig:AlfInt}. This leads to
particle heating as a result of the collisionless damping of the Alfv\'en
wave, also known as non-linear Landau damping.

Restricting ourselves to a setup of pairwise interacting waves, we can identify
two distinct cases. In the first case counter-propagating waves are
interacting. In consequence, both waves damp, lose energy to the longitudinal
wave and subsequently heat the particles.  In the second case the waves are
co-propagating.  Here the wave with the smaller wavelength will not only
transfer energy to the particles, but also to the other Alfv\'en wave.
\citet{leeDampingNonlinearWaveparticle1973} describe this mechanism in detail
and formulate the following coupled set of differential equations while adopting
a measure for the magnetic energy of a wave, $I_j = \abs{B_j}^2$, where
$j\in\{1, 2\}$:
\begin{equation}
\frac{\rm{d}}{\rm{d}t} I_j = 2 \Gamma_j I_j.
\end{equation}
The coupling between the differential equations is implicit because the damping
coefficient has the dependency $\Gamma_1 \propto I_2$.  For the
counter-propagating case with an isothermal ion-electron-plasma in the high beta
limit $\beta_{\rm i} = 2\mu_0 n_{\rm i} \kB T_{\rm i}/B_0^2 = 2 \gg 1$, where
$B_0$ is the background magnetic field strength, the damping rate $\Gamma_j$ is
approximately equal for both wave polarizations with similar frequencies $\omega_j$
and may be approximated by \citep{holcombMicrophysicsGyroresonantStreaming2019}
\begin{equation}
\Gamma_1 = - \dfrac{\sqrt{\upi}}{16} \dfrac{I_2}{B_0^2} \sqrt{\beta_{\rm i}} \omega_1.
\end{equation}
Note that $\Gamma_2$ is found by substituting the subscripts
$1\rightarrow2$ and $2\rightarrow1$.

\begin{figure}
	\centering
	\includegraphics{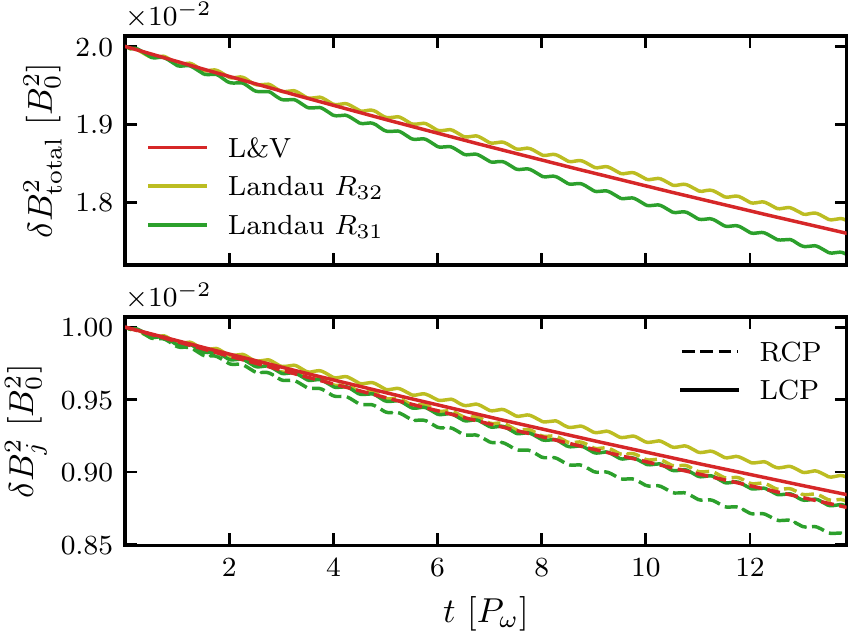}
	\caption{Time evolution of the magnetic energy of a linearly polarized
          Alfv\'en wave in our fluid simulations with Landau damping. Time is
          measured in units of the period of the mean wave frequencies $P_\omega
          = 4\upi(\omega_1 + \omega_2)^{-1}$. Analytical predictions for the
          damping rate are taken from \citet[][labelled
            L\&V]{leeDampingNonlinearWaveparticle1973}. The fluid simulations
          are presented with the different heat flux closures $R_{31}$ and
          $R_{32}$.  We compare the time evolution of the total magnetic wave
          energy (top panel) and the magnetic wave energy of the different
          polarization states (bottom panel).  The right-hand circularly polarized
          wave has a higher phase velocity and loses energy more quickly
          in comparison to the left-hand circularly polarized wave.}
	\label{fig:LinBeta2}
\end{figure}

In figure~\ref{fig:LinBeta2} we show simulations of a linearly polarized Alfv\'en wave, which consists of two counter-propagating waves of equal amplitude.
The pure fluid simulations are shown with a box size of $L=252\,c/\omega_{\rm i}$ and wavelengths $\uplambda=L/3$. Right and left polarized waves are initialized with phase velocities $\omega_{\rm RCP}/k = 0.0342$ and $\omega_{\rm LCP}/k = 0.0318$ with a perpendicular magnetic field amplitude of $\delta B = 0.1\, B_0$.
A reduced mass-ratio of $m_{\rm i}/m_{\rm e} =100$ is adapted here. 

Our simulations are carried out with the different heat flux closures $R_{32}$
and $R_{31}$, as shown in figure~\ref{fig:LinBeta2}.  Both closures reproduce the
theoretical predictions quite well.  A PIC simulation with similar parameters
has been shown in figure~6.4 by
\citet{holcombMicrophysicsGyroresonantStreaming2019}, which reproduces half of
the predicted damping rate until $t\sim2P_\omega$ and shows a quenching of the
damping rate afterwards. In comparison to kinetic simulations, there is no
saturation of the Landau-damping effect in fluids.  This is because the
distribution of the fluid particles is always assumed to be roughly Maxwellian
and resonant particles are not depleted as a function of time.  Hence, Landau fluid is
implicitly assumed to have small thermalization timescale in comparison to the
damping timescale. On the other hand, PIC simulations are plagued by Poisson
noise and an insufficient resolution of velocity space might lead to a reduced
Landau damping rate.

\subsection{Gyrotropic CR streaming instability}
\label{sec:CSRI}
\begin{figure}
\centering
\includegraphics[width=\textwidth]{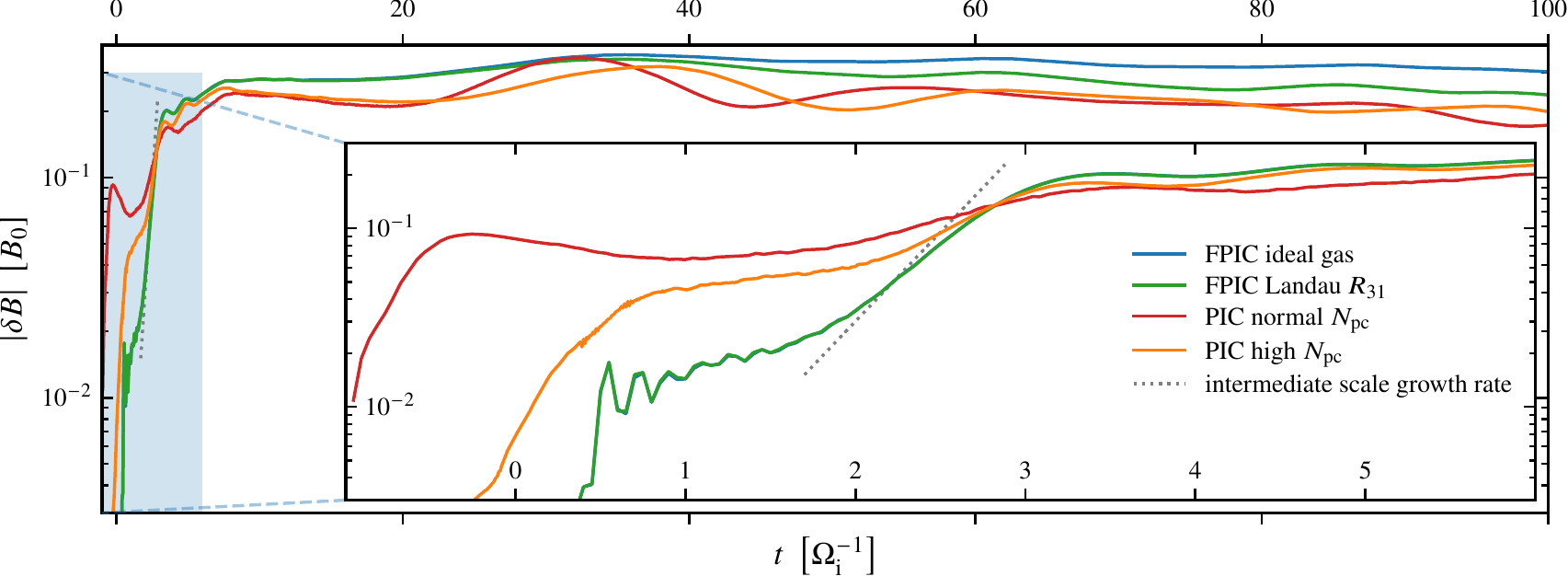}
\caption{Growth of the perpendicular magnetic field as a function of time for a
  gyrotropic CR streaming setup.  The maximum growth rate expected from
  the linear dispersion relation at intermediate scales is
  $\Gamma_{\mathrm{inter}} = 2.299 \Omega_{\rm i}$ and shown in dashed
  grey. because of the different initial seed populations for the particle
  species, the onset of the instabilities is not expected to happen at the same
  simulation time.Hence, we choose an arbitrary $t=0$ so that the different
  simulated growth phases roughly coincide.}
\label{fig:crGrowth}
	\centering
	\includegraphics[width=\textwidth]{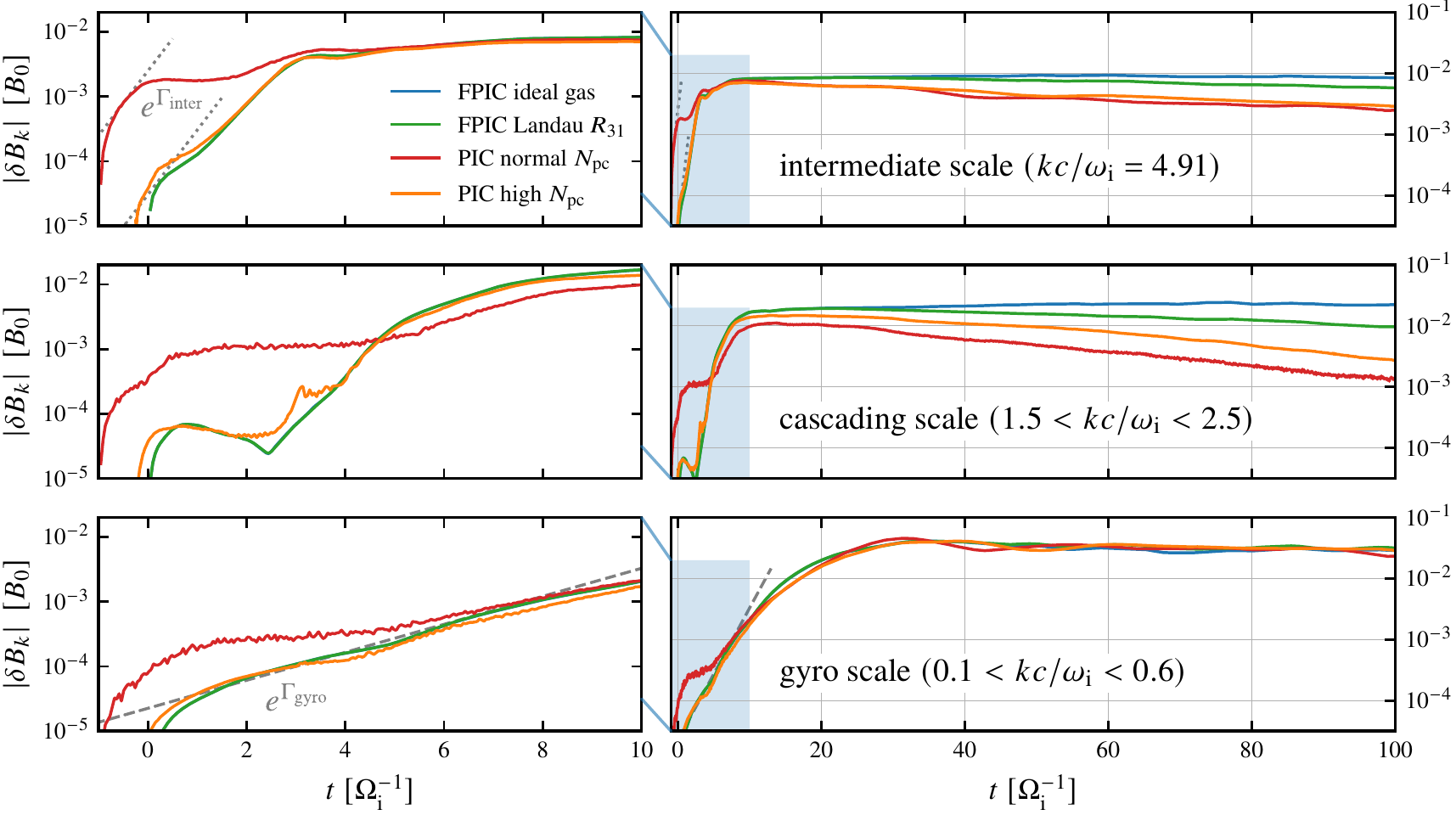}
	\caption{Growth of the perpendicular magnetic field as a function of time at
		different scales for a gyrotropic CR streaming setup. We show mean
		values of the fields that are averaged over a range of wave vectors $k$, as
		indicated in the legends. The maximum growth rates at the gyro scale and the
		intermediate scale are given by $\Gamma_{\mathrm{gyro}} = 0.498 \Omega_{\rm
			i}$ and $\Gamma_{\mathrm{inter}} = 2.299 \Omega_{\rm i}$, and indicated by
		the grey dotted and dashed lines, respectively.  At wavenumbers corresponding
		to cascading scales, there is no instability expected according to the linear
		dispersion relation, and wave growth solely arises as a result of cascading
		from other (unstable) scales.  }
	\label{fig:crScales}
\end{figure}

To test the entire code, we run CR streaming instability simulations, where electron and ion CRs are modelled with the PIC method and the background electron and ion plasmas are modelled as fluids.
The initial CR momentum distribution for ions (electrons) is assumed to be a gyrotropic distribution with a non-vanishing (zero) pitch angle, while both CR electrons and ions are assumed to drift at the same velocity $\varv_{\rm dr}$. 
Namely, the phase space distributions for the electron and ion CR species $s\in\{\mathrm{e, i}\}$ are given by \citep{shalabyNewCosmicRaydriven2020}
\begin{equation}
	f_{\mathrm{cr,} s}(\vec{x},\vec{u})
	=  \frac{ n_{\mathrm{cr,} s}  }{2\upi u_\perp}
	\delta(u_{\parallel} - \gamma_s \varv_{\rm dr}) \delta(u_{\perp}-\gamma_s \varv_{\perp, s}) ,
\end{equation}
where $\gamma_s = (1- \varv^2_{\rm dr} /c^2- \varv^2_{\perp, s}/c^2)^{-1/2}$ is  the Lorentz factor and $\varv_{\perp, s}$ is the perpendicular component of the CR velocity. We choose $\varv_{\perp,\mathrm e} = 0$ and $\varv_{\perp, \mathrm{i}}=13.1 \varv_{\rm A}$, where the ion Alfv\'en velocity is given by $\varv_{\rm A} = B_0/(\mu_0 n_{\rm i} m_{\rm i})^{1/2} = 0.01c$ with the background magnetic field pointing along the spatial direction, and $\varv_{\rm dr}$ of $5 \varv_{\rm A}$ resulting in a pitch angle for the ions of $\tan^{-1} (\varv_{\perp,\mathrm i}/\varv_{\rm dr} )= 69.1^{\circ}$. The thermal background species are isothermal with the temperatures $k_\mathrm{B} T/(m c^2)=10^{-4}$ and a mass ratio $m_{\rm i}/m_{\rm e} = 1836$. We use a periodic box of length $L_x = 10~971.5\,c/\omega_\mathrm{p}$ and resolution $\Delta x = 0.1\,c/\omega_\mathrm{p}$.
The CR to background ratio number density ratio $\alpha=n_{\rm cr, i}/n_{\rm i}=0.01$.

We run two simulations where the background plasmas are modelled as fluids.
The first one uses an ideal gas closure without accounting for Landau damping (FPIC ideal gas) while we include the heat flux source term in the second simulation to mimic the impact of linear Landau damping using the $R_{31}$ closure of equation~\eqref{eq:R31} (FPIC Landau $R_{31}$).
We compare these two fluid-PIC simulations against PIC simulations where both CRs and background plasmas are modelled as PIC species. 
The number of CR ions per cell is $N_{\mathrm{pc}} = 25~(75)$ and we call this simulation \enquote{PIC normal (high) $N_{\mathrm{pc}}$}~\citep{shalabyNewCosmicRaydriven2020}. Like the \enquote{PIC normal $N_{\mathrm{pc}}$} simulation, the fluid-PIC simulations also use $25$ particles per cell for modelling CRs.

Growth rates of the instability in the linear regime can be computed from the linear cold background plasma dispersion relation \citep{holcombGrowthSaturationGyroresonant2019,Shalaby2022}:
\begin{align}
	0 = &
	1 - \frac{k^2c^2}{\omega ^2}
	+ \frac{\omega _{\rm i}^2}{\omega  \left(-\omega  \pm \Omega _{\mathrm{i},0}\right)} +
	\frac{\omega _{\rm e}^2}{\omega  \left(-\omega \pm \Omega _{\mathrm{e},0}  \right)}
	+ \frac{\alpha  \omega _{\rm e}^2 }{ \gamma _{\rm e}  \omega ^2 }
	\left(
	\frac{ \omega -k \varv_{\mathrm{dr}} }
	{  k \varv_{\mathrm{dr}}-\omega \pm \Omega _{\mathrm{e},0}}
	\right)
	\nonumber \\ &
	+ \frac{\alpha  \omega _{\rm i}^2 }{ \gamma _{\rm i}  \omega ^2 }
	\left(
	\frac{ \omega -k \varv_{\mathrm{dr}} }
	{  k \varv_{\mathrm{dr}}-\omega   \pm\Omega _{\rm i}}
	- \frac{ \varv_{\perp}^2/c^2  \left(k^2 c^2-\omega ^2\right)}
	{2  \left(k \varv_{\mathrm{dr}}-\omega \pm\Omega _{\rm i} \right){}^2}
	\right).
\end{align}
The non-relativistic and relativistic cyclotron frequencies of each species are given by $\Omega_{\mathrm{s},0}=q_s B_0/m_s$ and $\Omega_{\mathrm{ s}}=\Omega_{\mathrm{s},0}/\gamma_\mathrm{s}$ respectively.
The wavelength of the most unstable wave mode at the gyroscale is $\uplambda_{\mathrm{g}} = 2\pi(\varv_{\rm dr}-\varv_{\rm A})/\Omega_{\rm i}$, which is properly captured in our setup using a box size of $L_x \sim 10.15 \uplambda_{\mathrm{g}}$.

We show the amplification of the perpendicular magnetic field components as a function of time for this unstable setup in figure~\ref{fig:crGrowth} for various simulations. 
It shows that the noise level of the fluid-PIC simulations is orders of magnitude lower in comparison to the \enquote{PIC normal $N_{\mathrm{pc}}$} resolution, even though the number of CR particles per cell is the same.
Especially up to the saturation point ($t\Omega_{\rm i}\sim 10$) the fluid-PIC simulation compares more favourably to the PIC results with lower noise than to the PIC simulation with fewer $N_{\mathrm{pc}}$.

After saturation, i.e.\ when Alfv\'en waves at many scales have built up and their interaction has created an electrostatic field, these waves start to lose some energy to Landau damping of the electrostatic waves (see section~\ref{sec:Alfven_waves}).
At that point, the Landau closure becomes relevant.
Qualitatively the ideal gas closure has no efficient mechanism for dissipating such electrostatic waves, resulting in a prolonged growth period leading to saturation at higher values at the cascading and intermediate scales.
Utilization of a Landau closure leads to some damping, albeit it is quantitatively smaller than in the PIC simulations.
While figure~\ref{fig:dp1f} indicates faster damping for the Landau closures in comparison to the kinetic results in the electron electrostatic branches, damping in the ion-acoustic branch might be underestimated in the Landau closures. We have compared the expected damping between kinetic and Landau fluid in the ion-acoustic branch for multiple wavenumbers, which confirmed that this is a likely scenario.
The accuracy of this approximation is not the same at all scales, which can be seen in figure~\ref{fig:crScales}, where the magnetic field amplifications at various ranges of scales are compared. Especially in the highly Landau-damped scales, differences between fluid-PIC and PIC emerge.
At ion gyro scales, where most of the magnetic energy is stored at saturation, there is a good agreement over the entire time period. 
Exponential growth at every scale is also in good agreement between PIC and fluid-PIC simulations at all scales.
The initial exponential growth can also be compared to the expected growth rates from the linear dispersion relation.
The growth rates of the two local maxima are plotted alongside the simulated data, one at the intermediate scales around $c k=4.91\omega_{\rm i}$ and one at the gyro scale at $c k = 0.38 \omega_{\rm i}$.
The intermediate scale starts an inverse cascade to larger scales almost immediately, which causes a reduced growth rate in comparison to the expectation from linear theory. By contrast, the gyro scale instability follows linear expectations to very good approximation.

While our fluid-PIC and PIC results are promisingly similar, differences after the saturation level might be attributed to multiple reasons. 
First, the Landau closures do not exactly reproduce the correct damping, and therefore will deviate quantitatively. 
Second, due to the high electron temperature chosen, relativistic effects might occur in PIC, but not in the non-relativistic fluid that we assumed for the background plasma. 
Third, the PIC method might exhibit more numerical dissipation at the given $N_{\rm{pc}}$ in comparison to the fluid method.
However, figure~\ref{fig:crScales} seems to indicate numerical convergence at the intermediate and gyro scale. 

Even though our simulations were run at unrealistically high $\alpha$, the background particles did not deviate significantly from the Maxwellian distribution at the end of the simulation time.
This indicates, that a fluid description for background species is indeed a valid approach for this setup, especially for smaller, more realistic values of $\alpha$.

\subsection{Computational scaling}
\label{sec:scaling}
\begin{figure}
\centering
\includegraphics{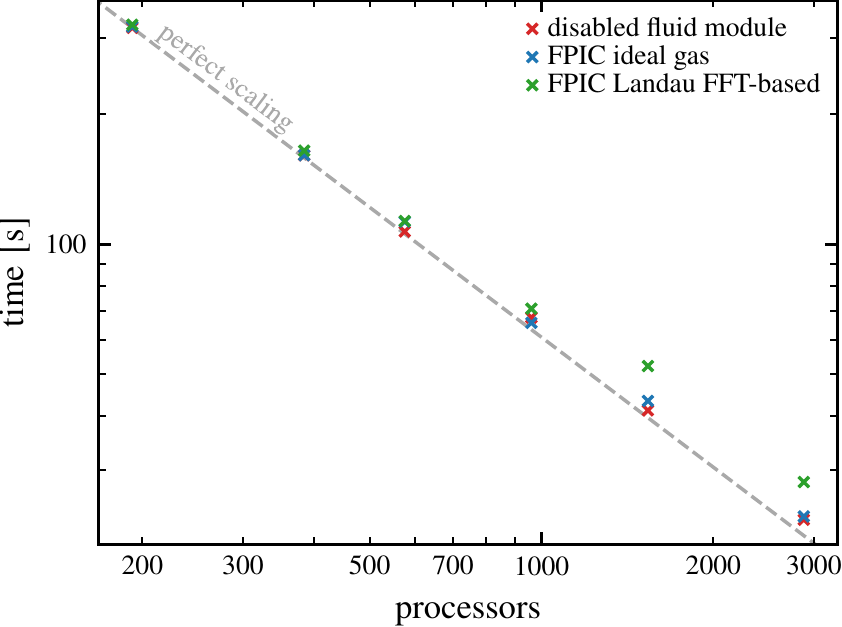}
\caption{Strong scaling of the fluid-PIC code, with and without Fourier-based
  Landau closures.  Shown is the wall-clock time needed to simulate $1250$ time
  integration steps with $180000$ cells at $1000$ particles per cell at a
  varying number of processors. We show the perfect strong scaling that is
  proportional to the inverse number of processors as the grey dashed line for
  reference.  For the disabled fluid module no background plasma was initialized
  and only CRs are initialized, showing that the bulk of the
  computational work is performed by the PIC routines.}
\label{fig:strong_scaling}
\end{figure}
We show the strong scaling properties of our fluid-PIC code in
Fig.~\ref{fig:strong_scaling}.  The tests were run on Intel Cascade 9242
processors with 96 processors per node at the HLRN Emmy cluster.  Simulations
with 3000 processors or more typically cause severe bottlenecks due to the
latency and/or the finite bandwidth of input/ouput operations. For this number
of processors the Fourier-based closures are roughly 20 per cent more costly
in comparison to the ideal gas closures. This is in stark contrast to pure PIC simulations,
which scale with the inverse ratio of CR-to-background density
$\alpha^{-1}$, consequently the fluid-PIC algorithm leads to a speed-up of a factor of $100$ for the simulation performed in section~\ref{sec:CSRI}, which adopted unrealistically large $\alpha$.

The bottleneck in the communication procedure of our implementation is currently
the \enquote{Ialltoallv} MPI routine, which is not optimized for hierarchical
architecture networks as of now. Further optimizations to this might provide
fruitful in increasing the code's scalability further if necessary.

The fluid-PIC simulations in section~\ref{sec:CSRI} used only $N_{\rm{pc}}=25$
and seem to be sufficiently resolved. For such a low particle number, the FFT is
the bottleneck for scalability because the overlap of communication and
computation is small, i.e.\ we measure a 260 per cent increase in time with 2880
processors, while at 192 processors the increase is below 20 per cent. This
indicates that scalability of fluid-only simulations is dominated quickly by the
FFT, while the cost is almost negligible for fluid-PIC simulations.  Still,
simulations with only a few particles per cell are computationally inexpensive
so that there is no reason for performing such a simulation on thousands of
processors. Furthermore, the example of a mono-energetic cold CR beam
is not very demanding regarding the phase-space resolution. More realistic
scenarios include power law distributions for the CR population as well as
larger spatial density inhomogeneities, both resulting in an increased
requirement for the number of particles per cell in order to accurately resolve
the velocity phase-space distribution along the entire spatial domain.

\section{Conclusion}
\label{sec:conclusion}
In this paper, we introduce a new technique termed fluid-PIC, which uses Maxwell's equations to self-consistently couple the PIC method to the fluid equations.
This technique is particularly aimed at simulating energetic particles like CRs interacting with a thermal plasma. This enables us to resolve effects on electron time and length scales and to emulate Landau damping in the fluid by incorporating appropriate closures for the divergence of the heat flux.
The underlying building blocks of our implementation are the SHARP 1D3V PIC-code extended by a newly developed fluid module and the overall algorithm is second-order accurate in space and time.
While an ideal fluid does not exhibit Landau damping, we have implemented two different Landau fluid closures and studied their performance. Here we summarize our main findings:
\begin{itemize}
\item We developed a \textit{stable} multi-species fluid code that is coupled to explicit PIC algorithm. In order to couple multi-fluid equations to Maxwell's equations, very often implicit and semi-implicit methods have been used for stability reasons. However, the resulting interdependency between all fluids complicates their coupling to explicit PIC methods. 
To ensure numerical stability, Riemann solvers that provide some numerical diffusion are used. However, we demonstrate that the level of numerical diffusivity needs to be carefully controlled so that it does not numerically damp small-amplitude plasma waves or quench plasma instabilities. Most importantly, our new fluid-PIC code fully resolves the electron timescales, precluding the need to adopt any simplifying assumptions to the electrical field components or to Ohm's law.

\item We compare various Landau fluid closures and demonstrate that local closures only produce reliable results close to a characteristic scale while they are prone to fail in multi-scale problems. By contrast, semi-local spatial filters or global (Fourier-based) methods to estimate Landau fluid closures produce reliable results for a large range of scales. Most importantly, we demonstrate that the inclusion of communication intensive (Fourier-based) fluid closures only have a minimal impact on our code performance (through the usage of non-blocking background communication) because the majority of the computational workload is taken up by the much more cost-intensive PIC module. This enables us to make use of the more accurate Fourier-based Landau closure for the fluid instead of relying on local approximations only.
  
\item In numerical tests, our implementation of the multi-species fluid module showed excellent agreement with theoretical frequencies and damping rates of Langmuir waves, oscillation frequencies of various two fluid wave modes, as well as the non-linear Landau damping of Alfv\'en waves.

\item First simulations of the CR streaming instability with our combined fluid-PIC code provide very good agreement with the results of pure PIC simulations, especially for the growth rates and saturation levels of the gyro-scale and intermediate-scale instabilities. This success is achieved at a substantially lower Poisson noise of the background plasma at the same number of computational CR particles per cell. Most importantly, the numerical cost of the fluid-PIC simulation is reduced by the CR-to-background number density ratio. However, we find that the late-time behaviour of the CR streaming instability differs for our fluid-PIC and PIC simulations. More work is needed to understand the reason for this, which could be either resulting from (i) numerical damping due to Poisson noise resulting from the finite number of PIC particles, (ii) missing relativistic (electron) effects in our non-relativistic fluid dynamics, or (iii) missing physics in our fluid closures that may be underestimating other relevant collisionless wave damping processes. 
  
\end{itemize}

Three possible future extensions of the algorithm are left open here.
(i) Extending the fluid formulation with a full pressure tensor, (ii) extending the code to two or three spatial dimensions, and (iii) the inclusion of direct interaction terms between the various fluids to explicitly incorporate scattering processes such as ion-neutral damping.
The novel fluid-PIC framework greatly extends the computationally limited parameter space accessible to pure PIC methods whilst not compromising on some of the most important microphysical plasma effects.
This opens up many possibilities for studying CR physics in physically relevant parameter regimes, such as the growth and saturation of the CR streaming instability in different environments, and including the effect of partial ionization, ion-neutral damping and inhomogeneities of the background plasma.

\section*{Acknowledgements}
The authors acknowledge support by the European Research Council under ERC-CoG grant CRAGSMAN-646955 and ERC-AdG grant PICOGAL-101019746. The work was supported by the North-German Supercomputing Alliance (HLRN), project bbp00046.

\section*{Data Availability}
The data underlying this article will be shared on reasonable request to the corresponding author.



\bibliographystyle{jpp}
\bibliography{citations}

\begin{thebibliography}{90}
\expandafter\ifx\csname natexlab\endcsname\relax\def\natexlab#1{#1}\fi
\def\au#1{#1} \def\ed#1{#1} \def\yr#1{#1}\def\at#1{#1}\def\jt#1{\textit{#1}}
  \def\bt#1{#1}\def\bvol#1{\textbf{#1}} \def\vol#1{#1} \def\pg#1{#1}
  \def\publ#1{#1}\def\arxiv#1{#1}\def\org#1{#1}\def\st#1{\textit{#1}}

\bibitem[{Allmann-Rahn} {\em et~al.\/}(2022){Allmann-Rahn}, Lautenbach \&
  Grauer]{allmann-rahnEnergyConservingVlasov2021}
{\sc \au{{Allmann-Rahn}, F.}, \au{Lautenbach, S.} \& \au{Grauer, R.}} \yr{2022}
   \at{An {{Energy Conserving Vlasov Solver That Tolerates Coarse Velocity
  Space Resolutions}}: {{Simulation}} of {{MMS Reconnection Events}}}.
  \jt{Journal of Geophysical Research: Space Physics}  \bvol{127}~(2),
  \pg{e2021JA029976}.

\bibitem[{Allmann-Rahn} {\em et~al.\/}(2018){Allmann-Rahn}, Trost \&
  Grauer]{allmann-rahnTemperatureGradientDriven2018}
{\sc \au{{Allmann-Rahn}, F.}, \au{Trost, T.} \& \au{Grauer, R.}} \yr{2018}
  \at{Temperature gradient driven heat flux closure in fluid simulations of
  collisionless reconnection}.  \jt{Journal of Plasma Physics}  \bvol{84}~(3).

\bibitem[Bai {\em et~al.\/}(2015)Bai, Caprioli, Sironi \&
  Spitkovsky]{baiMAGNETOHYDRODYNAMICPARTICLEINCELLMETHODCOUPLING2015}
{\sc \au{Bai, Xue-Ning}, \au{Caprioli, Damiano}, \au{Sironi, Lorenzo} \&
  \au{Spitkovsky, Anatoly}} \yr{2015}
  \at{{{MAGNETOHYDRODYNAMIC-PARTICLE-IN-CELL METHOD FOR COUPLING COSMIC RAYS
  WITH A THERMAL PLASMA}}: {{APPLICATION TO NON-RELATIVISTIC SHOCKS}}}.
  \jt{ApJ}  \bvol{809}~(1),  \pg{55}.

\bibitem[Bailey(1990)]{baileyFFTsExternalHierarchical1990}
{\sc \au{Bailey, David~H.}} \yr{1990}  \at{{{FFTs}} in external or hierarchical
  memory}.  \jt{J Supercomput}  \bvol{4}~(1),  \pg{23--35}.

\bibitem[Bell(2004)]{bellTurbulentAmplificationMagnetic2004}
{\sc \au{Bell, A.~R.}} \yr{2004}  \at{Turbulent amplification of magnetic field
  and diffusive shock acceleration of cosmic rays}.  \jt{\mnras}
  \bvol{353}~(2),  \pg{550--558}.

\bibitem[Birdsall \& Langdon(1991)]{birdsallPlasmaPhysicsComputer1991}
{\sc \au{Birdsall, Charles~K.} \& \au{Langdon, A.~Bruce}} \yr{1991} {\em Plasma
  Physics via Computer Simulation\/}.  \publ{{Bristol, Philadelphia}: {Adam
  Hilger ; IOP}}.

\bibitem[Blandford \&
  Eichler(1987)]{blandfordParticleAccelerationAstrophysical1987}
{\sc \au{Blandford, Roger} \& \au{Eichler, David}} \yr{1987}  \at{Particle
  acceleration at astrophysical shocks: {{A}} theory of cosmic ray origin}.
  \jt{Physics Reports}  \bvol{154}~(1),  \pg{1--75}.

\bibitem[Boris {\em et~al.\/}(1970)]{boris1970relativistic}
{\sc \au{Boris, Jay~P} \& \au{others}} \yr{1970} Relativistic plasma
  simulation-optimization of a hybrid code.  \bt{In {\em Proc. {{Fourth}} Conf.
  {{Num}}. {{Sim}}. {{Plasmas}}\/}},  \pg{pp. 3--67}.

\bibitem[Boulares \& Cox(1990)]{boularesGalacticHydrostaticEquilibrium1990}
{\sc \au{Boulares, Ahmed} \& \au{Cox, Donald~P.}} \yr{1990}  \at{Galactic
  {{Hydrostatic Equilibrium}} with {{Magnetic Tension}} and {{Cosmic-Ray
  Diffusion}}}.  \jt{The Astrophysical Journal}  \bvol{365},  \pg{544}.

\bibitem[Boyd \& Sanderson(2003)]{boydPhysicsPlasmas2003}
{\sc \au{Boyd, T. J.~M.} \& \au{Sanderson, J.~J.}} \yr{2003} {\em The
  {{Physics}} of {{Plasmas}}\/}.  \publ{{Cambridge}: {Cambridge University
  Press}}.

\bibitem[{Bret} \& {Dieckmann}(2010)]{Bret2010mime}
{\sc \au{{Bret}, A.} \& \au{{Dieckmann}, M.~E.}} \yr{2010}  \at{{How large can
  the electron to proton mass ratio be in particle-in-cell simulations of
  unstable systems?}}  \jt{Physics of Plasmas}  \bvol{17}~(3),  \pg{032109}.

\bibitem[{Bret} {\em et~al.\/}(2010){Bret}, {Gremillet} \&
  {Dieckmann}]{Bret2010}
{\sc \au{{Bret}, A.}, \au{{Gremillet}, L.} \& \au{{Dieckmann}, M.~E.}}
  \yr{2010}  \at{{Multidimensional electron beam-plasma instabilities in the
  relativistic regime}}.  \jt{Physics of Plasmas}  \bvol{17}~(12),
  \pg{120501}.

\bibitem[Buck {\em et~al.\/}(2020)Buck, Pfrommer, Pakmor, Grand \&
  Springel]{buckEffectsCosmicRays2020}
{\sc \au{Buck, Tobias}, \au{Pfrommer, Christoph}, \au{Pakmor, R{\"u}diger},
  \au{Grand, Robert J.~J.} \& \au{Springel, Volker}} \yr{2020}  \at{The effects
  of cosmic rays on the formation of {{Milky Way-mass}} galaxies in a
  cosmological context}.  \jt{Monthly Notices of the Royal Astronomical
  Society}  \bvol{497}~(2),  \pg{1712--1737}.

\bibitem[Burrows {\em et~al.\/}(2014)Burrows, Ao \&
  Zank]{burrowsNewHybridMethod2014}
{\sc \au{Burrows, R.~H.}, \au{Ao, X.} \& \au{Zank, G.~P.}} \yr{2014} A new
  hybrid method.  \bt{In {\em Outstanding Problems in Heliophysics: {{From}}
  Coronal Heating to the Edge of the Heliosphere\/} (ed. \ed{Q.~Hu \& G.~P.
  Zank})},  \st{Astronomical Society of the Pacific Conference Series},
  \vol{vol. 484},  \pg{p.~8}.

\bibitem[Butcher(2016)]{butcherNumericalMethodsOrdinary2016}
{\sc \au{Butcher, J.~C.}} \yr{2016} {\em Numerical {{Methods}} for {{Ordinary
  Differential Equations}}\/}, 3rd edn.  \publ{{Wiley}}.

\bibitem[Capdeville(2008)]{2008Capdeville}
{\sc \au{Capdeville, G.}} \yr{2008}  \at{A central {{WENO}} scheme for solving
  hyperbolic conservation laws on non-uniform meshes}.  \jt{Journal of
  Computational Physics}  \bvol{227}~(5),  \pg{2977--3014}.

\bibitem[Caprioli \& Spitkovsky(2014{\natexlab{{\em
  a\/}}})]{caprioliSimulationsIonAcceleration2014a}
{\sc \au{Caprioli, D.} \& \au{Spitkovsky, A.}} \yr{2014{\natexlab{{\em a\/}}}}
  \at{Simulations of {{Ion Acceleration}} at {{Non-relativistic Shocks}}.
  {{II}}. {{Magnetic Field Amplification}}}.  \jt{The Astrophysical Journal}
  \bvol{794},  \pg{46}.

\bibitem[Caprioli \& Spitkovsky(2014{\natexlab{{\em
  b\/}}})]{caprioliSimulationsIonAcceleration2014b}
{\sc \au{Caprioli, D.} \& \au{Spitkovsky, A.}} \yr{2014{\natexlab{{\em b\/}}}}
  \at{Simulations of {{Ion Acceleration}} at {{Non-relativistic Shocks}}.
  {{III}}. {{Particle Diffusion}}}.  \jt{The Astrophysical Journal}
  \bvol{794},  \pg{47}.

\bibitem[Cooley \& Tukey(1965)]{cooleyAlgorithmMachineCalculation1965}
{\sc \au{Cooley, James~W.} \& \au{Tukey, John~W.}} \yr{1965}  \at{An algorithm
  for the machine calculation of complex {{Fourier}} series}.  \jt{Math. Comp.}
   \bvol{19}~(90),  \pg{297--301}.

\bibitem[Cravero {\em et~al.\/}(2018{\natexlab{{\em a\/}}})Cravero, Puppo,
  Semplice \& Visconti]{2017Cravero}
{\sc \au{Cravero, I.}, \au{Puppo, G.}, \au{Semplice, M.} \& \au{Visconti, G.}}
  \yr{2018{\natexlab{{\em a\/}}}}  \at{Cool {{WENO}} schemes}.  \jt{Computers
  \& Fluids}  \bvol{169},  \pg{71--86}.

\bibitem[Cravero {\em et~al.\/}(2018{\natexlab{{\em b\/}}})Cravero, Puppo,
  Semplice \& Visconti]{craveroCWENOUniformlyAccurate2018}
{\sc \au{Cravero, I.}, \au{Puppo, G.}, \au{Semplice, M.} \& \au{Visconti, G.}}
  \yr{2018{\natexlab{{\em b\/}}}}  \at{{{CWENO}}: {{Uniformly}} accurate
  reconstructions for balance laws}.  \jt{Mathematics of Computation}
  \bvol{87}~(312),  \pg{1689--1719}.

\bibitem[Dalgarno(2006)]{dalgarnoInterstellarChemistrySpecial2006}
{\sc \au{Dalgarno, A.}} \yr{2006}  \at{Interstellar {{Chemistry Special
  Feature}}: {{The}} galactic cosmic ray ionization rate}.  \jt{Proceedings of
  the National Academy of Science}  \bvol{103},  \pg{12269--12273}.

\bibitem[{Daughton} {\em et~al.\/}(2011){Daughton}, {Roytershteyn},
  {Karimabadi}, {Yin}, {Albright}, {Bergen} \& {Bowers}]{Daughton2011}
{\sc \au{{Daughton}, W.}, \au{{Roytershteyn}, V.}, \au{{Karimabadi}, H.},
  \au{{Yin}, L.}, \au{{Albright}, B.~J.}, \au{{Bergen}, B.} \& \au{{Bowers},
  K.~J.}} \yr{2011}  \at{{Role of electron physics in the development of
  turbulent magnetic reconnection in collisionless plasmas}}.  \jt{Nature
  Physics}  \bvol{7}~(7),  \pg{539--542}.

\bibitem[{Daughton} {\em et~al.\/}(2006){Daughton}, {Scudder} \&
  {Karimabadi}]{Daughton2006}
{\sc \au{{Daughton}, William}, \au{{Scudder}, Jack} \& \au{{Karimabadi}, Homa}}
  \yr{2006}  \at{{Fully kinetic simulations of undriven magnetic reconnection
  with open boundary conditions}}.  \jt{Physics of Plasmas}  \bvol{13}~(7),
  \pg{072101}.

\bibitem[Dawson(1962)]{dawsonOneDimensionalPlasmaModel1962}
{\sc \au{Dawson, John}} \yr{1962}  \at{One-{{Dimensional Plasma Model}}}.
  \jt{Phys. Fluids}  \bvol{5}~(4),  \pg{445}.

\bibitem[Dimits {\em et~al.\/}(2014)Dimits, Joseph \&
  Umansky]{dimitsFastNonFourierMethod2014}
{\sc \au{Dimits, A.~M.}, \au{Joseph, I.} \& \au{Umansky, M.~V.}} \yr{2014}
  \at{A fast non-{{Fourier}} method for {{Landau-fluid}} operators}.
  \jt{Physics of Plasmas}  \bvol{21}~(5),  \pg{055907}.

\bibitem[Ding {\em et~al.\/}(2015)Ding, Li \&
  Chen]{dingHighorderAlgorithmsRiesz2015}
{\sc \au{Ding, Hengfei}, \au{Li, Changpin} \& \au{Chen, YangQuan}} \yr{2015}
  \at{High-order algorithms for {{Riesz}} derivative and their applications
  ({{II}})}.  \jt{Journal of Computational Physics}  \bvol{293},
  \pg{218--237}.

\bibitem[Draine(2011)]{drainePhysicsInterstellarIntergalactic2011}
{\sc \au{Draine, Bruce~T.}} \yr{2011} {\em Physics of the Interstellar and
  Intergalactic Medium\/}.  \publ{{Princeton, N.J}: {Princeton University
  Press}}.

\bibitem[Einfeldt {\em et~al.\/}(1991)Einfeldt, Munz, Roe \&
  Sj{\"o}green]{einfeldtGodunovtypeMethodsLow1991}
{\sc \au{Einfeldt, B}, \au{Munz, C.~D}, \au{Roe, P.~L} \& \au{Sj{\"o}green, B}}
  \yr{1991}  \at{On {{Godunov-type}} methods near low densities}.  \jt{Journal
  of Computational Physics}  \bvol{92}~(2),  \pg{273--295}.

\bibitem[Farber {\em et~al.\/}(2018)Farber, Ruszkowski, Yang \&
  Zweibel]{farberImpactCosmicRay2018}
{\sc \au{Farber, R.}, \au{Ruszkowski, M.}, \au{Yang, H.-Y.~K.} \& \au{Zweibel,
  E.~G.}} \yr{2018}  \at{Impact of {{Cosmic Ray Transport}} on {{Galactic
  Winds}}}.  \jt{ApJ}  \bvol{856}~(2),  \pg{112}.

\bibitem[Gargat{\'e} {\em et~al.\/}(2007)Gargat{\'e}, Bingham, Fonseca \&
  Silva]{gargateDHybridMassivelyParallel2007}
{\sc \au{Gargat{\'e}, L.}, \au{Bingham, R.}, \au{Fonseca, R.~A.} \& \au{Silva,
  L.~O.}} \yr{2007}  \at{{{dHybrid}}: {{A}} massively parallel code for hybrid
  simulations of space plasmas}.  \jt{Computer Physics Communications}
  \bvol{176}~(6),  \pg{419--425}.

\bibitem[Girichidis {\em et~al.\/}(2018)Girichidis, Naab, Hanasz \&
  Walch]{girichidisCoolerSmootherImpact2018}
{\sc \au{Girichidis, Philipp}, \au{Naab, Thorsten}, \au{Hanasz, Micha{\l}} \&
  \au{Walch, Stefanie}} \yr{2018}  \at{Cooler and smoother - the impact of
  cosmic rays on the phase structure of galactic outflows}.  \jt{Monthly
  Notices of the Royal Astronomical Society}  \bvol{479},  \pg{3042--3067}.

\bibitem[{Guo} \& {Oh}(2008)]{guo2008}
{\sc \au{{Guo}, Fulai} \& \au{{Oh}, S.~Peng}} \yr{2008}  \at{{Feedback heating
  by cosmic rays in clusters of galaxies}}.  \jt{\mnras}  \bvol{384}~(1),
  \pg{251--266}.

\bibitem[Hakim {\em et~al.\/}(2006)Hakim, Loverich \&
  Shumlak]{hakimHighResolutionWave2006}
{\sc \au{Hakim, A.}, \au{Loverich, J.} \& \au{Shumlak, U.}} \yr{2006}  \at{A
  high resolution wave propagation scheme for ideal {{Two-Fluid}} plasma
  equations}.  \jt{Journal of Computational Physics}  \bvol{219}~(1),
  \pg{418--442}.

\bibitem[Hammett \& Perkins(1990)]{hammettFluidMomentModels1990}
{\sc \au{Hammett, Gregory~W.} \& \au{Perkins, Francis~W.}} \yr{1990}  \at{Fluid
  moment models for {{Landau}} damping with application to the
  ion-temperature-gradient instability}.  \jt{Phys. Rev. Lett.}
  \bvol{64}~(25),  \pg{3019--3022}.

\bibitem[{Hanasz} {\em et~al.\/}(2013){Hanasz}, {Lesch}, {Naab}, {Gawryszczak},
  {Kowalik} \& {W{\'o}lta{\'n}ski}]{hanasz2013}
{\sc \au{{Hanasz}, M.}, \au{{Lesch}, H.}, \au{{Naab}, T.}, \au{{Gawryszczak},
  A.}, \au{{Kowalik}, K.} \& \au{{W{\'o}lta{\'n}ski}, D.}} \yr{2013}
  \at{{Cosmic Rays Can Drive Strong Outflows from Gas-rich High-redshift Disk
  Galaxies}}.  \jt{ApJL}  \bvol{777}~(2),  \pg{L38}.

\bibitem[Harten \& Hyman(1983)]{hartenSelfAdjustingGrid1983}
{\sc \au{Harten, Ami} \& \au{Hyman, James~M}} \yr{1983}  \at{Self adjusting
  grid methods for one-dimensional hyperbolic conservation laws}.  \jt{Journal
  of Computational Physics}  \bvol{50}~(2),  \pg{235--269}.

\bibitem[Hockney(1988)]{hockneyComputerSimulationUsing1988}
{\sc \au{Hockney, Roger~W.}} \yr{1988} {\em Computer Simulation Using
  Particles\/}.  \publ{{CRC Press}}.

\bibitem[Holcomb \& Spitkovsky(2019)]{holcombGrowthSaturationGyroresonant2019}
{\sc \au{Holcomb, Cole} \& \au{Spitkovsky, Anatoly}} \yr{2019}  \at{On the
  {{Growth}} and {{Saturation}} of the {{Gyroresonant Streaming
  Instabilities}}}.  \jt{ApJ}  \bvol{882}~(1),  \pg{3}.

\bibitem[Holcomb(2019)]{holcombMicrophysicsGyroresonantStreaming2019}
{\sc \au{Holcomb, Cole~James}} \yr{2019}  \at{The {{Microphysics}} of
  {{Gyroresonant Streaming Instabilities}} and {{Cosmic Ray
  Self-Confinement}}}. PhD thesis, Princeton University.

\bibitem[{Hong} {\em et~al.\/}(2012){Hong}, {Lee}, {Min} \& {Parks}]{Hong+2012}
{\sc \au{{Hong}, Jinhy}, \au{{Lee}, Ensang}, \au{{Min}, Kyoungwook} \&
  \au{{Parks}, George~K.}} \yr{2012}  \at{{Effect of ion-to-electron mass ratio
  on the evolution of ion beam driven instability in particle-in-cell
  simulations}}.  \jt{Physics of Plasmas}  \bvol{19}~(9),  \pg{092111}.

\bibitem[Hunana {\em et~al.\/}(2019)Hunana, Tenerani, Zank, Goldstein, Webb,
  Khomenko, Collados, Cally, Adhikari \&
  Velli]{hunanaIntroductoryGuideFluid2019}
{\sc \au{Hunana, P.}, \au{Tenerani, A.}, \au{Zank, G.~P.}, \au{Goldstein,
  M.~L.}, \au{Webb, G.~M.}, \au{Khomenko, E.}, \au{Collados, M.}, \au{Cally,
  P.~S.}, \au{Adhikari, L.} \& \au{Velli, M.}} \yr{2019}  \at{An introductory
  guide to fluid models with anisotropic temperatures. {{Part}} 2. {{Kinetic}}
  theory, {{Pad\'e}} approximants and {{Landau}} fluid closures}.  \jt{Journal
  of Plasma Physics}  \bvol{85}~(6).

\bibitem[Hunana {\em et~al.\/}(2018)Hunana, Zank, Laurenza, Tenerani, Webb,
  Goldstein, Velli \& Adhikari]{hunanaNewClosuresMore2018}
{\sc \au{Hunana, P.}, \au{Zank, G.~P.}, \au{Laurenza, M.}, \au{Tenerani, A.},
  \au{Webb, G.~M.}, \au{Goldstein, M.~L.}, \au{Velli, M.} \& \au{Adhikari, L.}}
  \yr{2018}  \at{New {{Closures}} for {{More Precise Modeling}} of {{Landau
  Damping}} in the {{Fluid Framework}}}.  \jt{Phys. Rev. Lett.}
  \bvol{121}~(13),  \pg{135101}.

\bibitem[Jacob \& Pfrommer(2017)]{jacobCosmicRayHeating2017}
{\sc \au{Jacob, Svenja} \& \au{Pfrommer, Christoph}} \yr{2017}  \at{Cosmic ray
  heating in cool core clusters \textendash{} {{I}}. {{Diversity}} of steady
  state solutions}.  \jt{Monthly Notices of the Royal Astronomical Society}
  \bvol{467}~(2),  \pg{1449--1477}.

\bibitem[Ji {\em et~al.\/}(2020)Ji, Chan, Hummels, Hopkins, Stern, Kere{\v s},
  Quataert, {Faucher-Gigu{\`e}re} \&
  Murray]{jiPropertiesCircumgalacticMedium2020}
{\sc \au{Ji, Suoqing}, \au{Chan, T~K}, \au{Hummels, Cameron~B}, \au{Hopkins,
  Philip~F}, \au{Stern, Jonathan}, \au{Kere{\v s}, Du{\v s}an}, \au{Quataert,
  Eliot}, \au{{Faucher-Gigu{\`e}re}, Claude-Andr{\'e}} \& \au{Murray, Norman}}
  \yr{2020}  \at{Properties of the circumgalactic medium in cosmic
  ray-dominated galaxy haloes}.  \jt{\mnras}  \bvol{496}~(4),  \pg{4221--4238}.

\bibitem[Jiang \& Shu(1996)]{1996Jiang}
{\sc \au{Jiang, Guang-Shan} \& \au{Shu, Chi-Wang}} \yr{1996}  \at{Efficient
  implementation of weighted {{ENO}} schemes}.  \jt{Journal of Computational
  Physics}  \bvol{126}~(1),  \pg{202--228}.

\bibitem[Kulsrud \& Pearce(1969)]{kulsrudEffectWaveParticleInteractions1969}
{\sc \au{Kulsrud, Russell} \& \au{Pearce, William~P.}} \yr{1969}  \at{The
  {{Effect}} of {{Wave-Particle Interactions}} on the {{Propagation}} of
  {{Cosmic Rays}}}.  \jt{ApJ}  \bvol{156},  \pg{445}.

\bibitem[Langdon \& Birdsall(1970)]{langdonTheoryPlasmaSimulation1970}
{\sc \au{Langdon, A.~Bruce} \& \au{Birdsall, Charles~K.}} \yr{1970}  \at{Theory
  of {{Plasma Simulation Using Finite}}-{{Size Particles}}}.  \jt{The Physics
  of Fluids}  \bvol{13}~(8),  \pg{2115--2122}.

\bibitem[Lee \& V{\"o}lk(1973)]{leeDampingNonlinearWaveparticle1973}
{\sc \au{Lee, Martin~A.} \& \au{V{\"o}lk, Heinrich~J.}} \yr{1973}  \at{Damping
  and nonlinear wave-particle interactions of {{Alfv\'en-waves}} in the solar
  wind}.  \jt{Astrophys Space Sci}  \bvol{24}~(1),  \pg{31--49}.

\bibitem[Lipatov(2002)]{lipatovHybridMultiscaleSimulation2002}
{\sc \au{Lipatov, Alexander~S.}} \yr{2002} {\em The {{Hybrid Multiscale
  Simulation Technology}}\/}.  \publ{{Berlin, Heidelberg}: {Springer Berlin
  Heidelberg}}.

\bibitem[{Marcowith} {\em et~al.\/}(2016){Marcowith}, {Bret}, {Bykov},
  {Dieckman}, {O'C Drury}, {Lemb{\`e}ge}, {Lemoine}, {Morlino}, {Murphy},
  {Pelletier}, {Plotnikov}, {Reville}, {Riquelme}, {Sironi} \& {Stockem
  Novo}]{Marcowith_2016-shock-review}
{\sc \au{{Marcowith}, A.}, \au{{Bret}, A.}, \au{{Bykov}, A.}, \au{{Dieckman},
  M.~E.}, \au{{O'C Drury}, L.}, \au{{Lemb{\`e}ge}, B.}, \au{{Lemoine}, M.},
  \au{{Morlino}, G.}, \au{{Murphy}, G.}, \au{{Pelletier}, G.}, \au{{Plotnikov},
  I.}, \au{{Reville}, B.}, \au{{Riquelme}, M.}, \au{{Sironi}, L.} \&
  \au{{Stockem Novo}, A.}} \yr{2016}  \at{{The microphysics of collisionless
  shock waves}}.  \jt{Reports on Progress in Physics}  \bvol{79}~(4),
  \pg{046901}.

\bibitem[Marcowith {\em et~al.\/}(2021)Marcowith, {van Marle} \&
  Plotnikov]{marcowithCosmicRaydrivenStreaming2021b}
{\sc \au{Marcowith, A.}, \au{{van Marle}, A.~J.} \& \au{Plotnikov, I.}}
  \yr{2021}  \at{The cosmic ray-driven streaming instability in astrophysical
  and space plasmas}.  \jt{Physics of Plasmas}  \bvol{28}~(8),  \pg{080601}.

\bibitem[{Moreno} {\em et~al.\/}(2018){Moreno}, {Dieckmann}, {Ribeyre},
  {Jequier}, {Tikhonchuk} \& {d'Humi{\`e}res}]{Moreno+2018}
{\sc \au{{Moreno}, Q.}, \au{{Dieckmann}, M.~E.}, \au{{Ribeyre}, X.},
  \au{{Jequier}, S.}, \au{{Tikhonchuk}, V.~T.} \& \au{{d'Humi{\`e}res}, E.}}
  \yr{2018}  \at{{Impact of the electron to ion mass ratio on unstable systems
  in particle-in-cell simulations}}.  \jt{Physics of Plasmas}  \bvol{25}~(6),
  \pg{062125}.

\bibitem[Ng {\em et~al.\/}(2020)Ng, Hakim, Wang \&
  Bhattacharjee]{ngImprovedTenmomentClosure2020}
{\sc \au{Ng, Jonathan}, \au{Hakim, A.}, \au{Wang, L.} \& \au{Bhattacharjee,
  A.}} \yr{2020}  \at{An improved ten-moment closure for reconnection and
  instabilities}.  \jt{Physics of Plasmas}  \bvol{27}~(8),  \pg{082106}.

\bibitem[Padovani {\em et~al.\/}(2020)Padovani, Ivlev, Galli, Offner, Indriolo,
  {Rodgers-Lee}, Marcowith, Girichidis, Bykov \&
  Kruijssen]{padovaniImpactLowEnergyCosmic2020}
{\sc \au{Padovani, Marco}, \au{Ivlev, Alexei~V.}, \au{Galli, Daniele},
  \au{Offner, Stella S.~R.}, \au{Indriolo, Nick}, \au{{Rodgers-Lee}, Donna},
  \au{Marcowith, Alexandre}, \au{Girichidis, Philipp}, \au{Bykov, Andrei~M.} \&
  \au{Kruijssen, J. M.~Diederik}} \yr{2020}  \at{Impact of {{Low-Energy Cosmic
  Rays}} on {{Star Formation}}}.  \jt{Space Science Reviews}  \bvol{216},
  \pg{29}.

\bibitem[Pakmor {\em et~al.\/}(2016)Pakmor, Pfrommer, Simpson \&
  Springel]{pakmorGalacticWindsDriven2016}
{\sc \au{Pakmor, R.}, \au{Pfrommer, C.}, \au{Simpson, C.~M.} \& \au{Springel,
  V.}} \yr{2016}  \at{Galactic {{Winds Driven}} by {{Isotropic}} and
  {{Anisotropic Cosmic-Ray Diffusion}} in {{Disk Galaxies}}}.  \jt{The
  Astrophysical Journal}  \bvol{824},  \pg{L30}.

\bibitem[{Park} {\em et~al.\/}(1992){Park}, {Parker}, {Biglari}, {Chance},
  {Chen}, {Cheng}, {Hahm}, {Lee}, {Kulsrud}, {Monticello}, {Sugiyama} \&
  {White}]{Park1992}
{\sc \au{{Park}, W.}, \au{{Parker}, S.}, \au{{Biglari}, H.}, \au{{Chance}, M.},
  \au{{Chen}, L.}, \au{{Cheng}, C.~Z.}, \au{{Hahm}, T.~S.}, \au{{Lee}, W.~W.},
  \au{{Kulsrud}, R.}, \au{{Monticello}, D.}, \au{{Sugiyama}, L.} \&
  \au{{White}, R.}} \yr{1992}  \at{{Three-dimensional hybrid
  gyrokinetic-magnetohydrodynamics simulation}}.  \jt{Physics of Fluids B}
  \bvol{4}~(7),  \pg{2033--2037}.

\bibitem[Passot {\em et~al.\/}(2014)Passot, Henri, Laveder \&
  Sulem]{passotFluidSimulationsIon2014}
{\sc \au{Passot, Thierry}, \au{Henri, Pierre}, \au{Laveder, Dimitri} \&
  \au{Sulem, Pierre-Louis}} \yr{2014}  \at{Fluid simulations of ion scale
  plasmas with weakly distorted magnetic fields}.  \jt{Eur. Phys. J. D}
  \bvol{68}~(7),  \pg{207}.

\bibitem[{Pfrommer}(2013)]{pfrommer2013}
{\sc \au{{Pfrommer}, Christoph}} \yr{2013}  \at{{Toward a Comprehensive Model
  for Feedback by Active Galactic Nuclei: New Insights from M87 Observations by
  LOFAR, Fermi, and H.E.S.S.}}  \jt{ApJL}  \bvol{779}~(1),  \pg{10}.

\bibitem[Riquelme \& Spitkovsky(2009)]{riquelmeNonlinearStudyBell2009}
{\sc \au{Riquelme, Mario~A.} \& \au{Spitkovsky, Anatoly}} \yr{2009}
  \at{Nonlinear {{Study}} of {{Bell}}'s {{Cosmic Ray Current-Driven
  Instability}}}.  \jt{The Astrophysical Journal}  \bvol{694},  \pg{626--642}.

\bibitem[Roe(1981)]{roeApproximateRiemannSolvers1981}
{\sc \au{Roe, P.~L}} \yr{1981}  \at{Approximate {{Riemann}} solvers, parameter
  vectors, and difference schemes}.  \jt{Journal of Computational Physics}
  \bvol{43}~(2),  \pg{357--372}.

\bibitem[Ruszkowski {\em et~al.\/}(2017)Ruszkowski, Yang \&
  Reynolds]{ruszkowskiCosmicRayFeedbackHeating2017}
{\sc \au{Ruszkowski, Mateusz}, \au{Yang, H.-Y.~Karen} \& \au{Reynolds,
  Christopher~S.}} \yr{2017}  \at{Cosmic-{{Ray Feedback Heating}} of the
  {{Intracluster Medium}}}.  \jt{ApJ}  \bvol{844}~(1),  \pg{13}.

\bibitem[{Ruszkowski} {\em et~al.\/}(2017){Ruszkowski}, {Yang} \&
  {Zweibel}]{ruszkowski2017}
{\sc \au{{Ruszkowski}, Mateusz}, \au{{Yang}, H. Y.~Karen} \& \au{{Zweibel},
  Ellen}} \yr{2017}  \at{{Global Simulations of Galactic Winds Including
  Cosmic-ray Streaming}}.  \jt{ApJL}  \bvol{834}~(2),  \pg{208}.

\bibitem[{Shalaby} {\em et~al.\/}(2017){Shalaby}, {Broderick}, {Chang},
  {Pfrommer}, {Lamberts} \& {Puchwein}]{resolution-paper}
{\sc \au{{Shalaby}, Mohamad}, \au{{Broderick}, Avery~E.}, \au{{Chang}, Philip},
  \au{{Pfrommer}, Christoph}, \au{{Lamberts}, Astrid} \& \au{{Puchwein},
  Ewald}} \yr{2017}  \at{{Importance of Resolving the Spectral Support of
  Beam-plasma Instabilities in Simulations}}.  \jt{ApJ}  \bvol{848}~(2),
  \pg{81}.

\bibitem[Shalaby {\em et~al.\/}(2017)Shalaby, Broderick, Chang, Pfrommer,
  Lamberts \& Puchwein]{shalabySHARPSpatiallyHigherorder2017}
{\sc \au{Shalaby, Mohamad}, \au{Broderick, Avery~E.}, \au{Chang, Philip},
  \au{Pfrommer, Christoph}, \au{Lamberts, Astrid} \& \au{Puchwein, Ewald}}
  \yr{2017}  \at{{{SHARP}}: {{A Spatially Higher-order}}, {{Relativistic
  Particle-in-cell Code}}}.  \jt{ApJ}  \bvol{841}~(1),  \pg{52}.

\bibitem[Shalaby {\em et~al.\/}(2018)Shalaby, Broderick, Chang, Pfrommer,
  Lamberts \& Puchwein]{shalabyGrowthBeamplasmaInstabilities2018}
{\sc \au{Shalaby, Mohamad}, \au{Broderick, Avery~E.}, \au{Chang, Philip},
  \au{Pfrommer, Christoph}, \au{Lamberts, Astrid} \& \au{Puchwein, Ewald}}
  \yr{2018}  \at{Growth of beam-plasma instabilities in the presence of
  background inhomogeneity}.  \jt{ApJ}  \bvol{859}~(1),  \pg{45}.

\bibitem[Shalaby {\em et~al.\/}(2020)Shalaby, Broderick, Chang, Pfrommer,
  Puchwein \& Lamberts]{shalabyGrowthLongitudinalBeam2020}
{\sc \au{Shalaby, Mohamad}, \au{Broderick, Avery~E.}, \au{Chang, Philip},
  \au{Pfrommer, Christoph}, \au{Puchwein, Ewald} \& \au{Lamberts, Astrid}}
  \yr{2020}  \at{The growth of the longitudinal beam\textendash plasma
  instability in the presence of an inhomogeneous background}.  \jt{Journal of
  Plasma Physics}  \bvol{86}~(2).

\bibitem[{Shalaby} {\em et~al.\/}(2022){Shalaby}, {Lemmerz}, {Thomas} \&
  {Pfrommer}]{Shalaby2022}
{\sc \au{{Shalaby}, Mohamad}, \au{{Lemmerz}, Rouven}, \au{{Thomas}, Timon} \&
  \au{{Pfrommer}, Christoph}} \yr{2022}  \at{{The Mechanism of Efficient
  Electron Acceleration at Parallel Nonrelativistic Shocks}}.  \jt{ApJ}
  \bvol{932}~(2),  \pg{86}.

\bibitem[Shalaby {\em et~al.\/}(2021)Shalaby, Thomas \&
  Pfrommer]{shalabyNewCosmicRaydriven2020}
{\sc \au{Shalaby, Mohamad}, \au{Thomas, Timon} \& \au{Pfrommer, Christoph}}
  \yr{2021}  \at{A {{New Cosmic-Ray-driven Instability}}}.  \jt{ApJ}
  \bvol{908}~(2),  \pg{206}.

\bibitem[Shumlak {\em et~al.\/}(2011)Shumlak, Lilly, Reddell, Sousa \&
  Srinivasan]{shumlakAdvancedPhysicsCalculations2011}
{\sc \au{Shumlak, U.}, \au{Lilly, R.}, \au{Reddell, N.}, \au{Sousa, E.} \&
  \au{Srinivasan, B.}} \yr{2011}  \at{Advanced physics calculations using a
  multi-fluid plasma model}.  \jt{Computer Physics Communications}
  \bvol{182}~(9),  \pg{1767--1770}.

\bibitem[Simpson {\em et~al.\/}(2016)Simpson, Pakmor, Marinacci, Pfrommer,
  Springel, Glover, Clark \& Smith]{simpsonRoleCosmicRayPressure2016}
{\sc \au{Simpson, Christine~M.}, \au{Pakmor, R{\"u}diger}, \au{Marinacci,
  Federico}, \au{Pfrommer, Christoph}, \au{Springel, Volker}, \au{Glover, Simon
  C.~O.}, \au{Clark, Paul~C.} \& \au{Smith, Rowan~J.}} \yr{2016}  \at{The
  {{Role}} of {{Cosmic-Ray Pressure}} in {{Accelerating Galactic Outflows}}}.
  \jt{The Astrophysical Journal}  \bvol{827},  \pg{L29}.

\bibitem[Sironi \&
  Spitkovsky(2014)]{sironiRelativisticReconnectionEfficient2014}
{\sc \au{Sironi, Lorenzo} \& \au{Spitkovsky, Anatoly}} \yr{2014}
  \at{Relativistic {{Reconnection}}: An {{Efficient Source}} of {{Non-Thermal
  Particles}}}.  \jt{ApJ}  \bvol{783}~(1),  \pg{L21}.

\bibitem[Soares~Frazao \& Zech(2002)]{soaresfrazaoUndularBoresSecondary2002}
{\sc \au{Soares~Frazao, Sandra} \& \au{Zech, Yves}} \yr{2002}  \at{Undular
  bores and secondary waves -{{Experiments}} and hybrid finite-volume
  modelling}.  \jt{Journal of Hydraulic Research}  \bvol{40}~(1),  \pg{33--43}.

\bibitem[Spitkovsky(2008)]{spitkovskyParticleAccelerationRelativistic2008}
{\sc \au{Spitkovsky, Anatoly}} \yr{2008}  \at{Particle {{Acceleration}} in
  {{Relativistic Collisionless Shocks}}: {{Fermi Process}} at {{Last}}?}
  \jt{ApJ}  \bvol{682}~(1),  \pg{L5}.

\bibitem[Stix(1992)]{stixWavesPlasmas1992}
{\sc \au{Stix, Thomas~Howard}} \yr{1992} {\em Waves in Plasmas\/}.  \publ{{New
  York}: {AIP}}.

\bibitem[Strang(1968)]{strangConstructionComparisonDifference1968}
{\sc \au{Strang, Gilbert}} \yr{1968}  \at{On the {{Construction}} and
  {{Comparison}} of {{Difference Schemes}}}.  \jt{SIAM J. Numer. Anal.}
  \bvol{5}~(3),  \pg{506--517}.

\bibitem[Takahashi \& Kanada(2000)]{takahashiHighPerformanceRadix2Parallel2000}
{\sc \au{Takahashi, Daisuke} \& \au{Kanada, Yasumasa}} \yr{2000}
  \at{High-{{Performance Radix-2}}, 3 and 5 {{Parallel}} 1-{{D Complex FFT
  Algorithms}} for {{Distributed-Memory Parallel Computers}}}.  \jt{The Journal
  of Supercomputing}  \bvol{15}~(2),  \pg{207--228}.

\bibitem[Thomas \& Pfrommer(2019)]{thomasCosmicrayHydrodynamicsAlfv2019}
{\sc \au{Thomas, Timon} \& \au{Pfrommer, Christoph}} \yr{2019}  \at{Cosmic-ray
  hydrodynamics: {{Alfv}}\textbackslash 'en-wave regulated transport of cosmic
  rays}.  \jt{Monthly Notices of the Royal Astronomical Society}
  \bvol{485}~(3),  \pg{2977--3008}.

\bibitem[Thomas {\em et~al.\/}(2020)Thomas, Pfrommer \&
  En{\ss}lin]{thomasProbingCosmicRay2020}
{\sc \au{Thomas, Timon}, \au{Pfrommer, Christoph} \& \au{En{\ss}lin,
  Torsten~A.}} \yr{2020}  \at{Probing {{Cosmic Ray Transport}} with {{Radio
  Synchrotron Harps}} in the {{Galactic Center}}}.  \jt{ApJ}  \bvol{890}~(2),
  \pg{L18}.

\bibitem[Thomas {\em et~al.\/}(2022)Thomas, Pfrommer \&
  Pakmor]{thomasCosmicRaydrivenGalactic2022}
{\sc \au{Thomas, Timon}, \au{Pfrommer, Christoph} \& \au{Pakmor, R{\"u}diger}}
  \yr{2022} Cosmic ray-driven galactic winds: Transport modes of cosmic rays
  and {{Alfv}}\textbackslash 'en-wave dark regions.

\bibitem[Toro(2009)]{toroRiemannSolversNumerical2009}
{\sc \au{Toro, E.~F.}} \yr{2009} {\em Riemann Solvers and Numerical Methods for
  Fluid Dynamics: A Practical Introduction\/}, 3rd edn.  \publ{{Dordrecht ; New
  York}: {Springer}}.

\bibitem[Toro {\em et~al.\/}(1994)Toro, Spruce \&
  Speares]{toroRestorationContactSurface1994}
{\sc \au{Toro, E.~F.}, \au{Spruce, M.} \& \au{Speares, W.}} \yr{1994}
  \at{Restoration of the contact surface in the {{HLL-Riemann}} solver}.
  \jt{Shock Waves}  \bvol{4}~(1),  \pg{25--34}.

\bibitem[Uhlig {\em et~al.\/}(2012)Uhlig, Pfrommer, Sharma, Nath, En{\ss}lin \&
  Springel]{uhligGalacticWindsDriven2012}
{\sc \au{Uhlig, M.}, \au{Pfrommer, C.}, \au{Sharma, M.}, \au{Nath, B.~B.},
  \au{En{\ss}lin, T.~A.} \& \au{Springel, V.}} \yr{2012}  \at{Galactic winds
  driven by cosmic ray streaming}.  \jt{Monthly Notices of the Royal
  Astronomical Society}  \bvol{423},  \pg{2374--2396}.

\bibitem[Umansky {\em et~al.\/}(2015)Umansky, Dimits, Joseph, Omotani \&
  Rognlien]{umanskyModelingTokamakDivertor2015}
{\sc \au{Umansky, M.~V.}, \au{Dimits, A.~M.}, \au{Joseph, I.}, \au{Omotani,
  J.~T.} \& \au{Rognlien, T.~D.}} \yr{2015}  \at{Modeling of tokamak divertor
  plasma for weakly collisional parallel electron transport}.  \jt{Journal of
  Nuclear Materials}  \bvol{463},  \pg{506--509}.

\bibitem[{van Marle} {\em et~al.\/}(2018){van Marle}, {Casse} \&
  {Marcowith}]{vanMarle2018}
{\sc \au{{van Marle}, Allard~Jan}, \au{{Casse}, Fabien} \& \au{{Marcowith},
  Alexandre}} \yr{2018}  \at{{On magnetic field amplification and particle
  acceleration near non-relativistic astrophysical shocks: particles in MHD
  cells simulations}}.  \jt{\mnras}  \bvol{473}~(3),  \pg{3394--3409}.

\bibitem[Wang {\em et~al.\/}(2015)Wang, Hakim, Bhattacharjee \&
  Germaschewski]{wangComparisonMultifluidMoment2015}
{\sc \au{Wang, Liang}, \au{Hakim, Ammar~H.}, \au{Bhattacharjee, A.} \&
  \au{Germaschewski, K.}} \yr{2015}  \at{Comparison of multi-fluid moment
  models with particle-in-cell simulations of collisionless magnetic
  reconnection}.  \jt{Physics of Plasmas}  \bvol{22}~(1),  \pg{012108}.

\bibitem[Wang {\em et~al.\/}(2020)Wang, Hakim, Ng, Dong \&
  Germaschewski]{wangExactLocallyImplicit2020}
{\sc \au{Wang, Liang}, \au{Hakim, Ammar~H.}, \au{Ng, Jonathan}, \au{Dong,
  Chuanfei} \& \au{Germaschewski, Kai}} \yr{2020}  \at{Exact and locally
  implicit source term solvers for multifluid-{{Maxwell}} systems}.
  \jt{Journal of Computational Physics}  \bvol{415},  \pg{109510}.

\bibitem[Wang {\em et~al.\/}(2019)Wang, Zhu, Xu \&
  Li]{wangLandaufluidClosureArbitrary2019}
{\sc \au{Wang, Libo}, \au{Zhu, Ben}, \au{Xu, Xue-qiao} \& \au{Li, Bo}}
  \yr{2019}  \at{A {{Landau-fluid}} closure for arbitrary frequency response}.
  \jt{AIP Advances}  \bvol{9}~(1),  \pg{015217}.

\bibitem[Xie(2014)]{xiePDRFGeneralDispersion2014}
{\sc \au{Xie, Hua-sheng}} \yr{2014}  \at{{{PDRF}}: {{A}} general dispersion
  relation solver for magnetized multi-fluid plasma}.  \jt{Computer Physics
  Communications}  \bvol{185}~(2),  \pg{670--675}.

\bibitem[Zweibel(2017)]{zweibelBasisCosmicRay2017}
{\sc \au{Zweibel, Ellen~G.}} \yr{2017}  \at{The basis for cosmic ray feedback:
  {{Written}} on the wind}.  \jt{Physics of Plasmas}  \bvol{24}~(5),
  \pg{055402}.

\end{thebibliography}



\appendix

\section{C-WENO coefficients}
\label{sec:cweno_coefficients}
We list all coefficients needed to implement the C-WENO reconstruction in this section. Because our reconstruction procedure is applied component-wise to each of the primitive variables, we assume for this appendix that we are reconstructing a single quantity $u$. The smoothness indicator for the low-order polynomials are given by \citep{1996Jiang}:
\begin{align}
  \textsc{IS}[P_\mathrm{L}] &= \frac{13}{12} \left(u_{i-2} - 2 u_{i-1} + u_{i} \right)^2 + \frac{1}{4} \left(u_{i-2} - 4 u_{i-1} + 3u_{i}\right)^2, \\
  \textsc{IS}[P_\mathrm{C}] &= \frac{13}{12} \left(u_{i-1} - 2 u_{i} + u_{i+1} \right)^2 + \frac{1}{4} \left(u_{i+1} - u_{i-1}\right)^2, \\
  \textsc{IS}[P_\mathrm{R}] &= \frac{13}{12} \left(u_{i} - 2 u_{i+1} + u_{i+2} \right)^2 + \frac{1}{4} \left(3u_{i} - 4 u_{i+1} + u_{i+2}\right)^2,
\end{align}
while four auxiliary variables are defined
\begin{align}
  D_1 &= \frac{\left(6 w_0-1\right) \left(u_{i-2}+u_{i+2}\right)-2 \left(18 w_0-1\right) \left(u_{i-1}-u_{i+1}\right)}{48 w_0}, \\
  D_2 &= \frac{\left(2 w_0-3\right) \left(u_{i-2}+u_{i+2}\right)-2 \left(2 w_0+9\right) u_i+ 12 \left(u_{i-1}+u_{i+1}\right)}{16 w_0},\\
  D_3 &= \frac{-u_{i-2}+2 \left(u_{i-1}-u_{i+1}\right)+u_{i+2}}{12 w_0},\\
  D_4 &= \frac{u_{i-2}-4 u_{i-1}+6 u_i-4 u_{i+1}+u_{i+2}}{24 w_0},
\end{align}
to define the smoothness indicator for the $P_0$ polynomial:
\begin{align}
  \textsc{IS}[P_0] &= D_1^2+\frac{13}{3}D_2^2+\frac{3129 }{80}D_3^2+\frac{87617}{140} D_4^2
  +\frac{1}{2}D_3 D_1+\frac{21}{5}  D_2 D_4.
\end{align}
The overall smoothness indicator is given by \citep{2017Cravero}:
\begin{equation}
  \tau = \left|\textsc{IS}[P_\mathrm{L}] - \textsc{IS}[P_\mathrm{R}]\right|.
\end{equation}

The low-order polynomials are evaluated at the left-hand interface of a given cell via:
\begin{align} 
P_\mathrm{L}\left(x_{i - \onehalf}\right) &= \frac{1}{6} (-u_{i-2} + 5u_{i-1} + 2 u_{i}), \\
P_\mathrm{C}\left(x_{i - \onehalf}\right) &= \frac{1}{6} (2u_{i-1} + 5u_{i} - u_{i+1}), \\
P_\mathrm{R}\left(x_{i - \onehalf}\right) &= \frac{1}{6} (11u_{i} - 7u_{i+1} + 2 u_{i+2}),
\end{align}
while they evaluate to
\begin{align} 
P_\mathrm{L}\left(x_{i + \onehalf}\right) &= \frac{1}{6} (2u_{i-2} - 7u_{i-1} + 11 u_{i}), \\
P_\mathrm{C}\left(x_{i + \onehalf}\right) &= \frac{1}{6} (-u_{i-1} + 5u_{i} + 2u_{i+1}), \\
P_\mathrm{R}\left(x_{i + \onehalf}\right) &= \frac{1}{6} (2u_{i} + 5u_{i+1} - u_{i+2}),
\end{align}
at the right-hand interface. The optimal polynomial evaluates to
\begin{align} 
P_\mathrm{opt}\left(x_{i - \onehalf}\right) &= \frac{1}{60} (-3u_{i-2} + 27u_{i-1} + 47u_{i} - 13u_{i+1} + 7u_{i+2}) \nonumber\\
&=\frac{1}{10} \left[ 3 P_\mathrm{L}\left(x_{i - \onehalf}\right) + 6 P_\mathrm{C}\left(x_{i - \onehalf}\right) +\phantom{1} P_\mathrm{R}\left(x_{i - \onehalf}\right) \right], \\
P_\mathrm{opt}\left(x_{i + \onehalf}\right) &= \frac{1}{10} \left[ \phantom{1} P_\mathrm{L}\left(x_{i + \onehalf}\right) + 6 P_\mathrm{C}\left(x_{i + \onehalf}\right) + 3 P_\mathrm{R}\left(x_{i + \onehalf}\right) \right],
\end{align}
at both interfaces of the cell. 
The interface values of $P_0$ can be derived from equation~\eqref{eq:P_0_definition}.

\section{Convergence order}
\label{sec:convergence}
\begin{figure}
	\centering
	\includegraphics{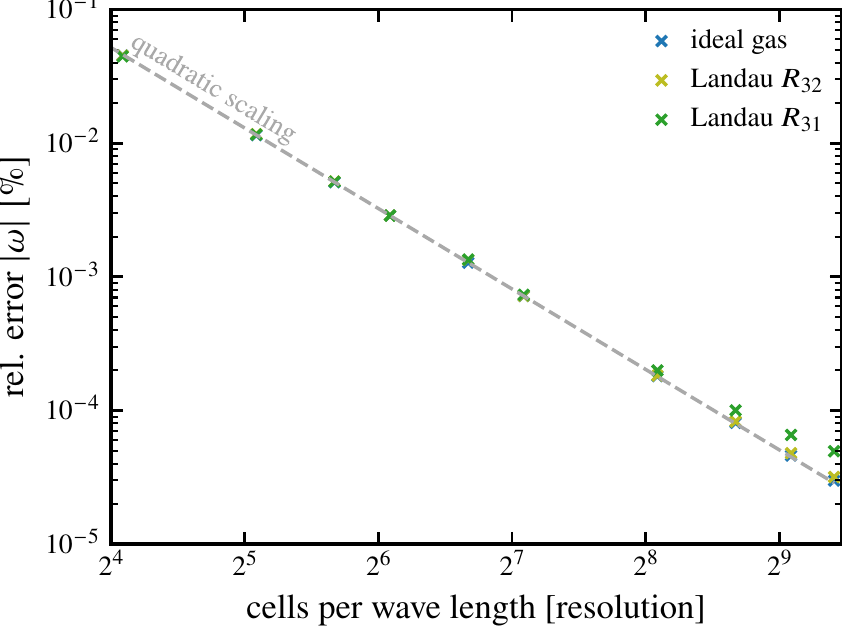}
	\caption{Relative error $\abs{(\omega^{\mathrm{sim}} -
			\omega^{\mathrm{theor}})/\omega^{\mathrm{theor}}}$ of the simulated frequency of a Langmuir wave at $k = 0.05 k_\mathrm{D}$.
		The same simulation setup is used in figure~\ref{fig:dp1f}, where we use a resolution of $68$ cells per wavelength.
		  The resolution here is varied between $68/4=17$ to $68\times10$ cells per wavelength. The grey line is a reference line for the second-order scaling of the error.}
	\label{fig:convergence}
\end{figure}
In order to numerically prove a second order scaling of the plasma frequency for the different heat flux closures, the linear dispersion of the Langmuir wave setup described in section~\ref{sec:Langmuir_damping} is simulated at different resolutions of $\uplambda/\Delta x$.
We  concentrate here on the convergence of a wave with wavenumber $k/k_\mathrm{D}=0.05$.
The results are shown in figure~\ref{fig:convergence} and demonstrate a very good match with the predicted errors assuming a second order convergence.
At first sight, the Landau closures do not seem to scale ideally for higher resolutions. However, this is the result of physical plasma heating due to wave damping in our setup leading to a non-linear increase in the expected plasma frequency.

\section{$R_{31}$ closure and adiabatic coefficients}
\label{sec:adiabatic}
While the $R_{32}$ closure assumes a fixed adiabatic index $\Gamma$ of $3$, the $R_{31}$ closure introduces a term proportional to $\hat\varw$ which alters the pressure equation in such a way that it increases the effective adiabatic index.   
To show this, we simplify equation~\eqref{eq:R31} by introducing the numerical coefficients $a_\varw$ and $a_T$ which are defined by comparing
\begin{equation}
\hat{Q} = a_\varw p_0 \hat{\varw} + \ci\,\mathrm{sign}\left(k\right) a_T \hat{T}.
\end{equation}
to equation~\eqref{eq:R31}. Using this ansatz and perturbing the pressure equation~\eqref{eq:pressure_update} with $p = p_0 + p_1$, where $p_1$ is the perturbation to the mean pressure $p_0$, in the absence of direct Landau damping ($a_T = 0$), we have
\begin{equation}
  \pdv{p_1}{t} = \left(-\Gamma p - a_\varw p_0 \right) \vec\nabla\bcdot\vec\varw
  -\vec\varw\bcdot\vec\nabla p 
  = \left(-\Gamma_\mathrm{eff} p_0 - \Gamma p_1 \right) \vec\nabla\bcdot\vec\varw
  -\vec\varw\bcdot\vec\nabla p,
\end{equation}
where $\Gamma_\mathrm{eff} = a_\varw + \Gamma = 4/(4-\upi) \simeq 4.66$ can be interpreted as the effective adiabatic index of the fluid. 
The evolution of sound waves of a non-electromagnetic fluid in the linear regime is governed by the linear term $\Gamma_\mathrm{eff} p_0 \vec\nabla\bcdot\vec\varw$ while the term $\Gamma p_1 \vec\nabla\bcdot\vec\varw$ adds non-linearity to this equation. In the linear approximation, the speed of sound becomes $c_\mathrm{s} = (\Gamma_\mathrm{eff}  p_0/ n_0)^{1/2}$ 
which coincides with the typical expression for the sound speed $c_\mathrm{s} = (\Gamma  p_0/ n_0)^{1/2}$ in the limit of $a_\varw=0$. 
This implies that the speed of sound is increased for the $R_{31}$ closure even if direct Landau damping is not present ($a_T = 0$). 
Interestingly, the effective adiabatic index and the speed of sound are independent of the choice of $\Gamma$. 
If direct Landau damping, as described by the $R_{31}$ closure, is affecting the fluid (i.e., $a_T \ne 0$), then the effective adiabatic index attains somewhat smaller values in comparison to $a_\varw + \Gamma$ while the wave frequency becomes complex because of the associated damping. Both are still independent of the choice of $\Gamma$.

This has consequences for simulations that model mildly relativistic fluids. If a simulation setup includes a fluid with an associated speed of sound near the speed of light $c_\mathrm{s} \lesssim c$, 
then a simulation that uses this setup with the $R_{31}$ closure can become unstable because $c_\mathrm{s}$ can now exceed the speed of light because of the aforementioned reason. 
\end{document}